\documentclass[twocolumn,showkeys,preprintnumbers,amsmath,amssymb]{revtex4-2}

\usepackage{graphicx}
\usepackage{dcolumn}
\usepackage{bm}
\usepackage{bbm}
\usepackage{amsmath}
\usepackage{mathtools}

\begin{document}

\title{Quantum Loads and the Generalized Telegrapher's Problem}

\author{E.$~$Collin$^{*,\dag}$ and A.$~$Delattre$^{*}$}

\address{(*) Univ. Grenoble Alpes, Institut N\'eel - CNRS UPR2940, 
25 rue des Martyrs, BP 166, 38042 Grenoble Cedex 9, France  }

\date{\today}

\begin{abstract}

We report on the mapping of classical microwave transmission line theory onto the quantum scattering matrix description. By means of a generalized flux formalism {\it a la Devoret}, we derive a charge-current vector living on the confining electrodes, to which {\it a Hodge-like decomposition} is applied.
The presented theory holds equally well {\it for all types of waves} traveling within Cartesian and cylindrical guide geometries (the usual Transverse Electric Magnetic, but also Transverse Magnetic and Transverse Electric ones).
We then demonstrate how a generic load decomposes into a series of complex coefficients $Z_{load}(\alpha)$, for each mode $\alpha$ propagating along the line. This decomposition 
leads to a particular classification of traveling waves: {\it gradient-type and curl-type} charge-currents. This distinction shines a new light on a {\it broken gauge symmetry} exhibited by 
a class of Transverse Electric modes. Our results 
provide a universal theoretical framework, which finds a direct usage in microwave-based quantum information transfer.
\end{abstract}


\maketitle

\section{Introduction}
\label{intro}

Condensed matter quantum information processing is based on microwave-operated devices installed in a milli-Kelvin cryogenic environment. 
The most emblematic of these components is certainly the quantum bit \cite{qubit}, the building block of quantum computers. 
But the quantum toolbox has been remarkably diversified over the last two decades, with for instance non-reciprocal elements based on opto-mechanics \cite{opto}, and traveling-wave quantum-limited amplifiers \cite{Twpa} (with no aim at exhaustiveness here). The quantum signals travel through {\it waveguides}, with a unique experimental implementation being the 30 meters long line used in a quantum bit entanglement demonstration reported in Ref. \cite{loopholewalraff}.

\begin{figure}[h!]
		\centering
	\includegraphics[width=10cm]{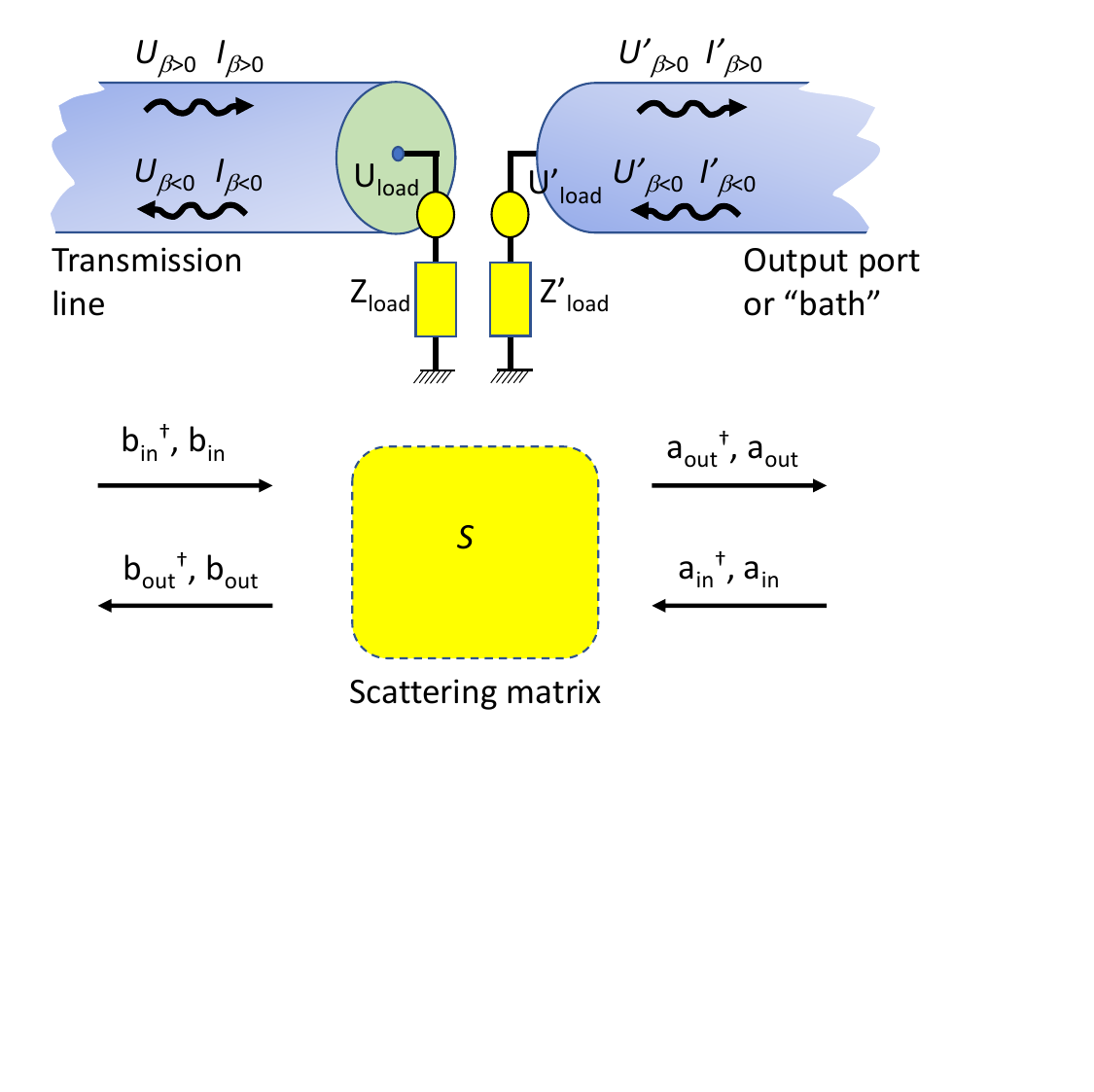}
           \vspace{-3.5cm}			
			\caption{ 
			Schematic of the problem at hand. Top: classical description of a transmission line which connects to another one (a "real waveguide", or a "pure electromagnetic bath"). Bottom: equivalent quantum formulation in terms of a scattering matrix $S$. }
			\label{fig1}
\end{figure}

The quantum description of traveling signals is based on the same grounds as {\it quantum optics: the widely recognized input-output theory} \cite {gardiner}. 
Specifically, photons voyage through lossless environments and then either interact with localized modes (like e.g. a microwave $LC$ resonator), or scatter at input/output ports. Rather recently, the finiteness of the photon wave-packet is being addressed theoretically with so-called {\it quantum pulses} \cite{molmer,quantumpulses}.  
Linear quantum amplifiers with their specific back-action properties have been modeled \cite{clerk}, 
and extensive review articles considering specifically  solid-state technologies can be found in the literature   
\cite{blencowe,nori,walraff,devoretRMP}.

Regarding the transport of quantum information in waveguides, the above mentioned conventional formalism is essential for the interpretation of measurements. However, it is far from complete: the quatization procedure is actually realized {\it only} for the Telegrapher's model Transverse Electric Magnetic field description \cite{pozar}, and the required scattering matrix which describes terminations is introduced in an {\it ad hoc} utilitarian fashion \cite{devoretRMP}.
The aim of the present manuscript is to go beyond these limitations, and provide an exact transcription of microwave engineering \cite{pozar} into quantum scattering \cite{devoretRMP}, see Fig. \ref{fig1}.
The presented full quantum modeling of wave transmission relies {\it directly} on the description of the fields in the most commonly used waveguides: square-type and cylinder-type. 
 This is performed with the introduction of a {\it generalized flux variable}, a quantity originally proposed by M. Devoret in the framework of quantum circuits \cite{devoret}.
The model rests on a strong differential geometry theorem: an {\it adapted Hodge's decomposition} \cite{frankel}. 

The final outcome is rather intuitive: a generic load is described by a {\it series of complex coefficients} $Z_{load}(\alpha)$, each one being characteristic of a given mode $\alpha$. The conventional formalism concerns only the mode-independent Transverse Electric Magnetic transport, for which a single $Z_{load}$ parameter is enough.  
All coefficients have strictly positive real part $Re(Z_{load}) >0$, and two extra phases emerge which enable to describe also non-reciprocal ports.
But Hodge's formalism brings in an important result: 
the traveling waves fall into two classes with {\it distinct geometrical properties}. And intriguingly, this must be related to the {\it gauge properties} of these electromagnetic modes.

\section{Starting point}
\label{start}

The reasoning developed here builds on a recent quantum modeling of Cartesian and cylindrical lossless waveguides, the simplest ones in terms of symmetries \cite{EddyguideNJP,AlexguideNJP}. In these two articles, we introduced a {\it generalized flux} \cite{devoret} defined on the confining electrodes:
\begin{eqnarray}
\!\!\!\varphi_\beta(x,z,t) \!\!& = &\!\! \phi_{zpf} \, g_\phi(x) \tilde{f}_\beta (z,t) , \label{varphi1} \\
\!\!\!\tilde{f}_\beta (z,t) \!\!& = &\!\!  (-\mathbbm{i}  b^\dag_\beta)\, e^{+\mathbbm{i}(\omega_\beta t- \beta z)} +(\mathbbm{i}  b_\beta)\, e^{-\mathbbm{i}(\omega_\beta t- \beta z)} \! ,
\end{eqnarray}
with $b^\dag_\beta$ ($b_\beta$) bosonic creation (annihilation) operators for a given propagating mode of wavevector $\beta \neq 0$, and corresponding angular frequency $\omega_\beta$ (we shall not use hat notations for operators in order to keep the writing lighter).
$\phi_{zpf}$ is a positive real number that carries the units, and will be given below. The normalized function $g_\phi(x)$ corresponds to the transverse profile of the $\varphi_\beta$ field, which propagates along $z$. 

\begin{figure}[h!]
		\centering
	\includegraphics[width=8cm]{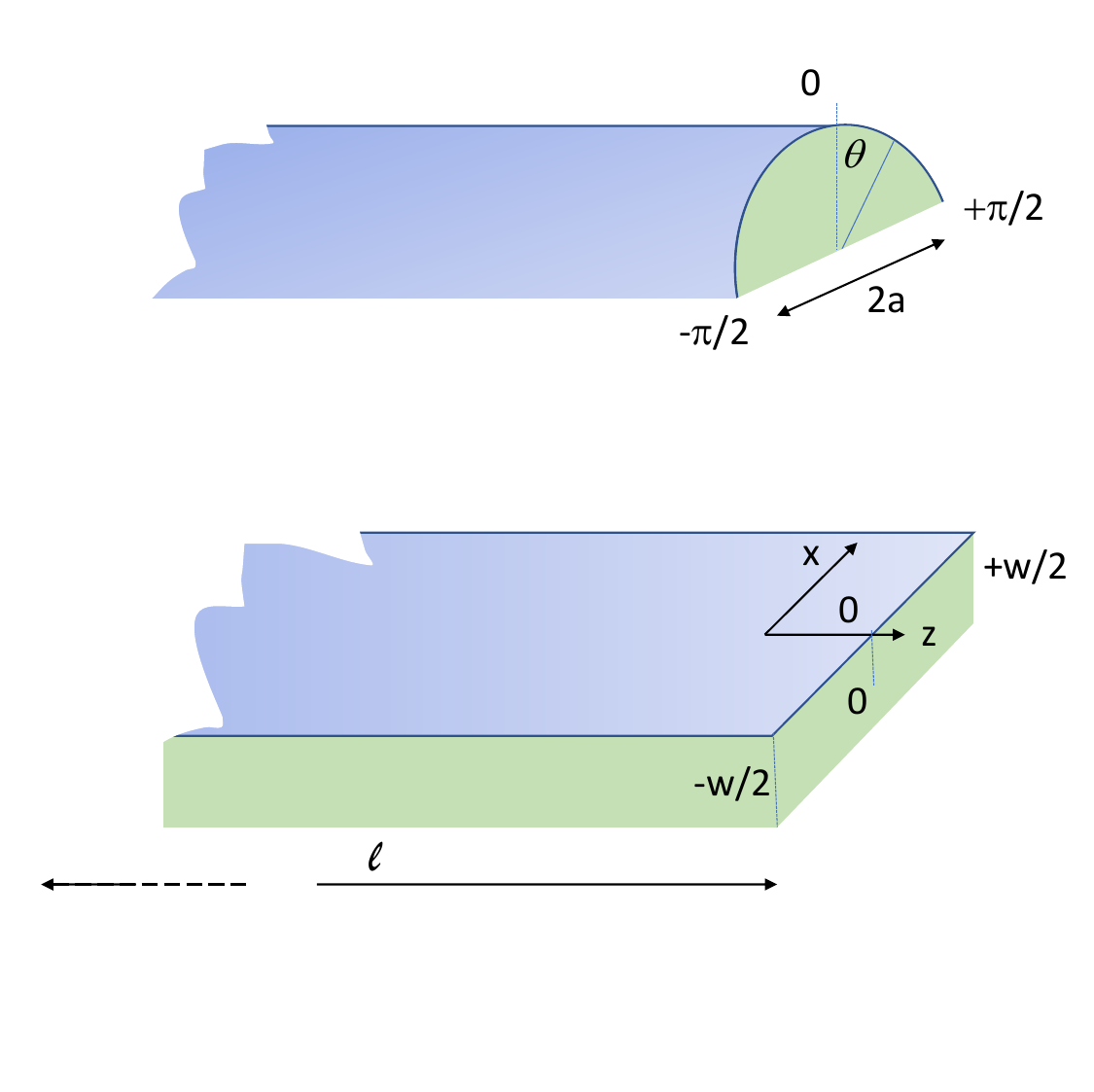}
	\vspace{-1.cm}
			\caption{
			One of the electrodes from the pair that confines microwaves in the envisaged guides. Top: the cylindrical (half-electrode) case \cite{AlexguideNJP}, which can be mapped onto the Cartesian (flat) case \cite{EddyguideNJP} (bottom). Metal in blue, dielectric in green. }
			\label{fig2}
\end{figure}

The configurations considered in Refs. \cite{EddyguideNJP,AlexguideNJP} are the parallel plates, the rectangular guide, the coaxial guide and the hollow cylinder. For the rectangular guide, two sets of equivalent electrodes face each other, and any of these pairs can be chosen to describe the problem at hand \cite{EddyguideNJP}. For the hollow cylinder, cutting the guide along a diameter we can define as well facing (half-) electrodes. 
Apart from the coaxial geometry, the electrodes are strictly identical; but in the latter case, the outer electrode is wider than the inner one. 
Note however that the modeling on each of them is strictly equivalent, and implies {\it the same} generalized flux field $\varphi_\beta$ \cite{AlexguideNJP}.
We shall thus discuss here only one of the electrodes of the pair. For a convenience that will be clarified below, in the coaxial case we always refer to the outer electrode.
These geometries are reminded in Fig. \ref{fig2}. 

Three families of waves exist in these guides \cite{pozar}: Transverse Electric Magnetic (TEM, only in parallel plates and coaxial), Transverse Magnetic (TM) and Transverse Electric (TE). 
For TM and TE, two indexes $\{n,m\}$ characterize 
a wave in addition to the wavevector $\beta$: they define {\it a branch} within the family. 
The symmetry of the parallel plate configuration is such that only $n=0$ is allowed; thus a single index $m$ is enough in this case.

In order to produce a uniform discussion that applies to all cases, we can conveniently write {\it for all configurations}:
\begin{eqnarray}
\omega_\beta & = & c \, k ,  \label{disperse}\\
k &=&\sqrt{\beta^2 + k_c^2} , \\ 
k_c^2 &= & k_{cx}^2+k_{c\eta}^2 , \label{kc}\\
k_{cx} &= & \frac{n \pi}{w}, \label{kcx}
\end{eqnarray}
with $c$ the speed of light in the guide and $w$ the width of the electrode.
For cylindrical geometries, the curved electrode is mapped onto a flat one with $x=a \theta$ the surface coordinate and $\theta \in [-\pi/2;\pi/2]$, see Fig. \ref{fig2}.
 $w= \pi a$ is then the half-perimeter, which is the adapted electrode size for the hollow cylinder. However for the coaxial line, a half is missing and one should simply use $2 w$ in the surface integral Eq. (\ref{weff}) below in order to recover the full-perimeter.
$k_c$ is the cutoff wavevector, which leads to a cutoff frequency $\omega_c = c \,k_c$ below which no TM,TE wave can propagate.
The cutoff wavevector splits in two contributions arising from each space coordinate: for a Cartesian symmetry $k_{c \eta}=k_{cy}$ while for cylindrical $k_{c \eta}=k_{cr}$ (which both depend on the $m$ index).
Eq. (\ref{kc}) obviously implies $k_{cx} \leq k_c$, 
 which is trivially satisfied for all but the coaxial case, where this is always true only if $w$ in Eq. (\ref{kcx}) corresponds to the outer electrode; this justifies our reference choice. 
By definition, we impose $k_{cx}=k_{c \eta}=0$ for the TEM waves (namely $n=m=0$): there is no cutoff and the dispersion relation Eq. (\ref{disperse}) is linear. 
All non-TEM propagating modes verify $m>0$ (for the TE ones in the rectangular waveguide, this amounts to a choice of facing electrodes since at least one integer of the $n,m$ pair must be nonzero in this case). Note that for TM waves in a rectangular microwave line, one has also $n>0$.

Considering all branches of the propagating families, {\it except the TE$_{n=0,m}$ ones}, we have:
\begin{eqnarray}
g_\phi(x) & = & \sin \left[ k_{cx} \left(x+\frac{w}{2}\right)\right] \,\, \mbox{for $n>0$}, \\
g_\phi(x) & = & 1 \,\,\,\, \mbox{for $n=0$}. 
\end{eqnarray}
Interestingly, the modeling leads to $g_\phi(x)=0$ for a  
TE$_{n=0,m}$ branch: there is {\it no} such generalized flux defined on the electrode, physically because of the absence of a charge and a longitudinal current density  \cite{EddyguideNJP,AlexguideNJP}. In the rectangular guide, a generalized flux is nonetheless present on the other pair of electrodes; but for any other configuration, 
one needs to invoke {\it virtual electrodes} in order to restore a propagating generalized flux field which shares the same properties as the one discussed here. 
These TE$_{n=0,m}$ configurations are actually quite special \cite{AlexguideNJP}, 
and require a specific discussion given in Appendix \ref{topomodes}; but our main conclusions still hold for them. We come back to those at the end of the manuscript in Section \ref{topodigr}, where {\it topology,  geometry, and gauge} are briefly addressed. \\

The propagation equations of the field $\varphi_\beta$ read \cite{EddyguideNJP,AlexguideNJP}:
\begin{eqnarray}
\frac{\partial^2 \varphi_\beta(x,z,t)}{\partial z^2}-\frac{1}{c^2} \frac{\partial^2 \varphi_\beta(x,z,t)}{\partial t^2} &=& 0 \,\, \mbox{for TEM}, \nonumber \\
\frac{\partial^2 \varphi_\beta(x,z,t)}{\partial z^2}-\frac{1}{v^2} \frac{\partial^2 \varphi_\beta(x,z,t)}{\partial t^2} &=& 0 \,\, \mbox{for TM}, \label{propag} \\
\frac{\partial^2 \varphi_\beta(x,z,t)}{\partial z^2}-\frac{1}{c^2} \frac{\partial^2 \varphi_\beta(x,z,t)}{\partial t^2} &=& k_c^2 \varphi_\beta(x,z,t) \,\, \mbox{for TE}, \nonumber
\end{eqnarray}
with $v= \omega_\beta /| \beta |$ the phase velocity.
While these expressions can be interpreted differently depending on the wave type (see Refs. \cite{EddyguideNJP,AlexguideNJP}), they can nonetheless be all recast in the same form: 
\begin{equation}
\frac{\partial^2 \varphi_\beta(x,z,t)}{\partial z^2}+\beta^2 \varphi_\beta(x,z,t)  = 0 . \label{genequa}
\end{equation} 
As well, charge and longitudinal current densities are defined from the generalized flux as:
\begin{eqnarray}
\sigma(x,z,t) & = & +C_z \frac{\partial  \varphi_\beta(x,z,t)}{\partial t } , \label{sigma} \\
j_z (x,z,t) & =& -L_z^{-1} \frac{\partial  \varphi_\beta(x,z,t)}{\partial z } \label{jz} .
\end{eqnarray}
By construction, $C_z=\epsilon/h_{ef\!f}$ for all wave types, and $L_z= \mu \, h_{ef\!f}$ for TEM and TE or $L_z= \mu \, h_{ef\!f} \,(\beta/k)^2$ for TM.
$\epsilon, \mu$ are the permittivity and permeability respectively of the dielectric present in the guide ($\epsilon \, \mu=1/c^2$), while $h_{ef\!f}$ is an effective thickness that can be computed from the geometry \cite{EddyguideNJP,AlexguideNJP}. 
Projecting the energy carried by the charge and current onto the transverse basis functions $g_\phi(x)$, one can define an effective electrode width:
\begin{equation}
w_{ef\!f} = \int_{-w/2}^{+w/2} g_\phi(x)^2 \, dx , \label{weff}
\end{equation}
with trivially $w_{ef\!f} = w$ for $n=0$ and $w_{ef\!f} = w/2$ for $n>0$. 
This enables to write:
\begin{eqnarray}
q(z,t) &=& w_{ef\!f} \, C_z \, \omega_\beta \,\phi_{zpf} f_\beta(z,t) , \label{charge} \\
I(z,t) & = & w_{ef\!f} \, \frac{\beta}{L_z } \, \phi_{zpf}  f_\beta(z,t) , \label{current} 
\end{eqnarray}
for the charge per unit length and current, with $f_\beta(z,t)$ the function in quadrature with $\tilde{f}_\beta(z,t)$:
\begin{equation}
\!\!\!\!\!\! f_\beta (z,t)   =   \mathbbm{i} \left[ (-\mathbbm{i}  b^\dag_\beta)\, e^{+\mathbbm{i}(\omega_\beta t- \beta z)} -(\mathbbm{i}  b_\beta)\, e^{-\mathbbm{i}(\omega_\beta t- \beta z)} \right] \! .
\end{equation}
Eqs. (\ref{charge},\ref{current}) 
have introduced the capacitance and inductance per unit length:
\begin{eqnarray}
C_\ell & = & w_{ef\!f} \, C_z  , \\
L_\ell & = & \frac{L_z}{w_{ef\!f}}. 
\end{eqnarray}
The total waveguide mode capacitance is obtained from the volume integral over the field stored energy \cite{EddyguideNJP,AlexguideNJP}:
\begin{eqnarray}
C_{tot} &  = &   C_z \, w_{ef\!f} \,  \ell \,\,\,\,\,\, \mbox{for TEM and TE} , \label{Ctot} \\
C_{tot} &  = &   C_z  \left(\frac{k}{\beta}\right)^{\!2}\!   w_{ef\!f} \,  \ell \,\,\,\,\,\, \mbox{for TM} , \label{Ctotbis} 
\end{eqnarray}
$\ell$ being the length of the waveguide (Fig. \ref{fig2}). 
The flux zero point fluctuation is finally given as:
\begin{equation}
\phi_{zpf}   =  \sqrt{\frac{\hbar}{2 \, C_{tot} \, \omega_\beta}} ,  
\end{equation}
with $\hbar$ Planck's reduced constant. \\

At the core of the quantum waveguide modeling of Refs. \cite{EddyguideNJP,AlexguideNJP} is the Devoret relationship \cite{devoret}:
\begin{equation}
\Delta V(x,z,t ) = \frac{ \partial \phi_\beta (x,z,t)}{\partial t} , 
\end{equation}
which links the generalized flux to an electrode voltage. This expression, which is actually borrowed from lumped elements circuit theory, implies here a {\it specific electromagnetic gauge choice} \cite{AlexguideNJP}. This is {\it not} a simple curiosity, and serves to interpret the specificity of TE$_{n=0,m}$ waves; see discussion in Section \ref{topodigr}. 
But for our purpose here, the point is that it enables us to define a voltage amplitude:
\begin{equation}
U(z,t) = \omega_\beta \, \phi_{zpf} f_\beta(z,t) , \label{U} 
\end{equation}
which is the voltage maximum in the transverse direction.
It corresponds to the classical quantity that propagates along with the current $I(z,t)$, as depicted in Fig. \ref{fig1}.
Injecting Eqs. (\ref{current},\ref{U}) into 
Eq. (\ref{genequa}), we obtain:
\begin{eqnarray}
\frac{\partial^2 U(z,t)}{\partial z^2} + \beta^2 \, U(z,t) & = & 0, \\
\frac{\partial^2 I(z,t)}{\partial z^2} + \beta^2 \, I(z,t) & = & 0,
\end{eqnarray}
which is the well-known {\it Telegrapher's problem} \cite{pozar}. 
From Eqs. (\ref{current},\ref{U}), we define then the transmission line {\it characteristic impedance}:
\begin{eqnarray}
Z_0 & = & \left| \frac{U(z,t)}{I(z,t)} \right|  \\
    & = & \frac{L_z \, \omega_\beta}{w_{ef\!f} \,|\beta|}. \label{caracZ}
\end{eqnarray}
With the wave impedance $Z_c = \sqrt{\mu/\epsilon}$ (characteristic of the medium),
we obtain for the different types of modes:
\begin{eqnarray}
Z_0 & = & \frac{h_{ef\!f} }{w_{ef\!f}} Z_c \,\,\,\, \mbox{for TEM}, \\
 Z_0 & = & \frac{h_{ef\!f} }{w_{ef\!f}} \left( \frac{ |\beta| }{k} \right) Z_c \,\,\,\, \mbox{for TM} , \\
 Z_0 & = & \frac{h_{ef\!f} }{w_{ef\!f}} \left( \frac{k }{ | \beta |} \right) Z_c \,\,\,\, \mbox{for TE} .
\end{eqnarray} 
The ratio $h_{ef\!f}/w_{ef\!f}$ is a mode-dependent geometrical parameter; only for TEM waves is $Z_0$ a  constant (meaning independent of $n, m$ and $\beta$). 
Note that the conventional expressions $v=1/\sqrt{C_\ell \, L_\ell}$ 
and $Z_0 =  \sqrt{L_\ell/C_\ell }$ are still valid for TEM and TM waves, but they {\it fail} in the case of TE modes. 
The mode-dependent parameters $h_{ef\!f}$, $w_{ef\!f}$ and $k_c$ are reminded in Appendix \ref{modeparams}. \\

This section has set up our framework, extending the conventional formalism that applies to TEM modes (see e.g. Ref. \cite{devoretRMP}) to the other families.
The aim of the following ones is to formalize 
{\it in what way} a termination affects the charges and currents that confine the electromagnetic field, 
{\it under which conditions} we can produce a scattering matrix description and finally {\it how} we can map it onto a 
classical model of a load impedance (see e.g. Ref. \cite{pozar}).

\section{The model}	
\label{model}

The generic solution describing the fields present in the waveguide is constructed from the superposition of all generalized flux fields Eqs. (\ref{varphi1}) propagating along the electrodes:
\begin{equation}
\varphi(x,z,t) = \!\!\!\sum_{\substack{ \mbox{TEM,TM,TE}^* \\  \{n,m\},\,\beta\,>0}} \!\!\! \varphi_\beta(x,z,t) + \varphi_{-\beta}(x,z,t), \label{varphi2}
\end{equation}
within the allowed branches. In the above, the $*$ means that we exclude again the TE$_{n =0,m}$ solutions which must be treated separately (Appendix \ref{topomodes}); the two equivalent waves propagating left and right are 
explicitly written such that from now $\beta >0$.
Knowing that the meaningful limit is $\ell \rightarrow +\infty$ (infinitely long waveguide), we chose periodic boundary conditions in $z$: $\beta= p \, 2\pi/\ell$, $p \in \mathbb{N}^*$ \cite{EddyguideNJP}.
Using a generic index $\alpha$ encompassing the branch type and $\beta$, Eq. (\ref{varphi2}) can be recast into:
\begin{eqnarray}
&& \varphi(x,z,t) = \sum_{\alpha } \varphi_\alpha(x,z,t) , \\
&&\varphi_\alpha(x,z,t) = \nonumber \\
&& - \sqrt{2} \, \mathbbm{i} \phi_{zpf} \,   g_\phi(x) \,  e^{+\mathbbm{i}\omega_\beta t} \! \left[ c^\dag_{s,\beta} \cos(\beta z)- \mathbbm{i}\,c^\dag_{a,\beta} \sin(\beta z) \right] \nonumber \\
&& + c.c. , \label{varphialpha}
\end{eqnarray}
with:
\begin{eqnarray}
c^\dag_{s,\beta} & =& \frac{b^\dag_{\beta}+b^\dag_{-\beta}}{\sqrt{2}}, \\
c^\dag_{a,\beta} & =& \frac{b^\dag_{\beta}-b^\dag_{-\beta}}{\sqrt{2}},
\end{eqnarray}
the symmetric and antisymmetric bosonic combinations (creation operators, and equivalently for annihilation ones; these obviously verify $[c_{i},c_{j}^\dag]= \delta_{ij}$, $[c_{i},c_{j}]=[c_{i}^\dag,c_{j}^\dag]=0$). \\

Eqs. (\ref{sigma},\ref{jz}) can be seen as components of a 2D+1 gradient field $\vec{j}_{grad}=\{j_{grad,x},j_{grad,z},j_{grad,t} \}$, having defined:
\begin{eqnarray}
t' & = & \mathbbm{i} \bar{v} \, t , \label{i1} \\
j_{grad,t} (x,z,t') & = & \mathbbm{i} \bar{v} \, \sigma(x,z,t) , \label{i2}  \\
\bar{v} & = & 1/\sqrt{C_z L_z} ,
\end{eqnarray}
which leads to, for each component $\alpha$:
\begin{eqnarray}
j_{grad,x} (x,z,t') & = & -\frac{1}{L_x} \frac{\partial  \varphi_\alpha (x,z,t)}{\partial x }, \\
j_{grad,z} (x,z,t') & = & -\frac{1}{L_z} \frac{\partial  \varphi_\alpha (x,z,t)}{\partial z }, \\
j_{grad,t} (x,z,t') & = & -\frac{1}{L_z} \frac{\partial  \varphi_\alpha (x,z,t)}{\partial t' } .
\end{eqnarray}
The convenient introduction of $\mathbbm{i} \bar{v}$ in Eqs. (\ref{i1},\ref{i2}) enables to symmetrize the current and charge equations.
As well, the subtlety here as compared to a "standard" gradient is that each axis is weighted by a  given inductance $L_i$, leading to the proper $j_{grad,i}$ ($i=x,z,t$). As such,
$j_{grad,x}$ is the transverse current present for TE$_{n\neq0,m}$ waves only;
 in this case, $L_x = L_z \, k_{cx}^2/k_{c}^2$. But for TEM and TM modes, we shall take $L_x \rightarrow +\infty$ at the end of the calculation and therefore always recover $j_{grad,x}=0$, as we should \cite{EddyguideNJP,AlexguideNJP}.

From the linearity of the problem at hand, the total 2D+1 charge-current vector $\vec{j}$ is simply the sum over $\alpha$ of all individual components. Consider one of those $ \vec{j}_\alpha$.
The reasoning is based on a "Hodge-like" theorem  applied to this vector, which construction is described in the present Section.
 The original theorem states (under assumptions clarified below) that $ \vec{j}_\alpha$ admits a unique decomposition into \cite{frankel}:
\begin{equation}
\vec{j}_\alpha = \vec{j}_{grad} + \vec{j}_{curl} + \vec{j}_{harm} , \label{hodgedecomp}
\end{equation}
where $\vec{j}_{grad}$ is our gradient vector given above, $\vec{j}_{curl}$ is a curl form and the harmonic term $\vec{j}_{harm}$ is neither of these two.
In the 2D+1 space we have defined, we pose for $\vec{j}_{curl}=\{ j_{curl,x},  j_{curl,z},  j_{curl,t} \}$:
\begin{eqnarray}
\!\!\!\!\!\!\!\!\!\!\!\!\!\!\!\! j_{curl,x} (x,z,t') & =&   \frac{1}{L_{x}} \frac{\partial \Psi_t(x,z,t)}{ \partial z} - \frac{1}{ L_{x}}  \frac{ \partial \Psi_z(x,z,t)}{ \partial t'}   , \label{curl1} \\
\!\!\!\!\!\!\!\!\!\!\!\!\!\!\!\! j_{curl,z} (x,z,t') & =&    \frac{1}{  L_{z}}\frac{\partial \Psi_x(x,z,t)}{ \partial t'} - \frac{1}{L_{x}} \frac{ \partial \Psi_t(x,z,t)}{ \partial x}   , \\
\!\!\!\!\!\!\!\!\!\!\!\!\!\!\!\! j_{curl,t} (x,z,t') & =&   \frac{1}{L_{x}}\frac{\partial \Psi_z(x,z,t)}{ \partial x} - \frac{1}{L_{z}} \frac{ \partial \Psi_x(x,z,t)}{ \partial z}    , \label{curl3}
\end{eqnarray}
with $ j_{curl,t} = \mathbbm{i} \bar{v} \,  \sigma_{curl}$ and $\vec{\Psi}_\alpha=\{ \Psi_x, \Psi_z, \Psi_t \}$ the curl vector associated to the component $\alpha$. Strictly speaking, this operator is a "generalized curl", in the sense that
we have again introduced weighting factors.
The associated divergence operator adopts the conventional writing:
\begin{equation}
\mbox{div} \left(\vec{j}\right) = \frac{\partial j_x}{\partial x} +\frac{\partial j_z}{\partial z}+\frac{\partial j_t}{\partial t'} ,
\end{equation}
for any vector field $\vec{j}$. Combining these definitions, one can come up with a "generalized Laplacian", formally of the type $\Delta_L=grad.div +L_x \, curl.curl$:
\begin{eqnarray}
j_{\Delta_L,x} & = & - \frac{1}{L_x} \frac{\partial^2 j_x}{\partial x^2}- \frac{1}{L_z} \frac{\partial^2 j_x}{\partial z^2} - \frac{1}{L_z} \frac{\partial^2 j_x}{\partial t'^2} , \nonumber \\
j_{\Delta_L,z} & = & - \frac{1}{L_x} \frac{\partial^2 j_z}{\partial x^2}- \frac{1}{L_z} \frac{\partial^2 j_z}{\partial z^2} - \frac{1}{L_z} \frac{\partial^2 j_z}{\partial t'^2}, \label{DeltaEq} \\
j_{\Delta_L,t} & = & - \frac{1}{L_x} \frac{\partial^2 j_t}{\partial x^2}- \frac{1}{L_z} \frac{\partial^2 j_t}{\partial z^2} - \frac{1}{L_z} \frac{\partial^2 j_t}{\partial t'^2},  \nonumber
\end{eqnarray}
having decomposed a generic $\Delta_L (\vec{j}\,)$.

 Our 
 "Hodge-like" mathematical 
 structure requires as well 
(symmetric, non-degenerate) bilinear forms on both the scalar field and the vector field which apply to mode $\alpha$. This is realized here with the weighted integrals:
\begin{eqnarray}
&& \langle \varphi_1,\varphi_2 \rangle_{scalar}  =  \nonumber \\ 
&&  \mathbbm{i} \bar{v}\!\! \int_{T} \int^0_{-\ell} \int_{-w}^{w} \varphi_1(x,z,t) \, \varphi_2(x,z,t) \, dtdzdx, \label{scal1} \\
&& \langle \vec{j}_1,\vec{j}_2 \rangle_{vector}  =  \nonumber \\
&&   \int_{T'} \int^0_{-\ell} \int_{-w}^{w} \left[ L_x \,  j_{1,x}(x,z,t')   \,j_{2,x}(x,z,t') \right. \nonumber \\
&& \left.\,\,\,\,\, + L_z \,  j_{1,z}(x,z,t')  \, j_{2,z}(x,z,t') \right. \nonumber \\
&& \left. + L_z \,  j_{1,t}(x,z,t')  \, j_{2,t}(x,z,t') \right] dt'dzdx ,\label{scal2}
\end{eqnarray}
with $T'=\mathbbm{i} \bar{v}\, T$ defined from 
the time period $T=2 \pi/\omega_\beta$, and (for now) the $z-$integral spanning the guide length defined between $-\ell$ and 0 (Fig. \ref{fig2}). Note however that the $x-$integral is taken over $2 w$, in order to span a full period of the function $g_\phi$ (even if outside of the electrode, it loses its physical interpretation). 
Also, both the inductances $L_x,L_z$ and the time period $T$ depend on the mode $\alpha$. Finally, the bilinear forms Eqs. (\ref{scal1},\ref{scal2}) are not positive definite, so these {\it are not} scalar products. 
This arises because in our fictitious 2D+1 space, our $\vec{j}$ charge-current operators {\it are not} especially Hermitian; this is the price we have to pay. 
Transforming back the fictitious fields into physical ones requires then a bit of caution with complex conjugates, as explained below with the curl $\vec{\Psi}_\alpha$ vector.
The absence of a proper scalar product will be briefly addressed in Section \ref{topodigr}. \\

It is easy to show that $\mbox{div}\,(\vec{j}_{curl})=0$ and that when applying the curl to our gradient $\vec{j}_{grad}$ instead of $\vec{\Psi}_\alpha$, one obtains zero as well. 
This actually means that our generalized gradient, curl and divergence operators define a {\it de Rham differential complex}. From Eqs. (\ref{scal1},\ref{scal2}) using proper {\it boundary conditions} (periodic, as discussed below), we can also demonstrate that  {\it gradient and divergence are adjoints, and that the curl is its own adjoint} \cite{frankel}.
As a consequence, gradient and curl forms are always orthogonal in the sense of Eq. (\ref{scal2}).
The harmonic part $\vec{j}_{harm}$ is then defined as 
$div\,(\vec{j}_{harm})=0$ together with zero curl. As a consequence:  $\Delta_L (\vec{j}_{harm})=0$, and $\vec{j}_{harm}$ is also orthogonal to gradients and curls; by definition $j_{harm,t}= \mathbbm{i} \bar{v} \,  \sigma_{harm}$.

Physically, $ \{\partial/\partial x,\partial/\partial z, \partial/\partial t'\} . \vec{j}_{\alpha} =0$ expresses  charge conservation \cite{EddyguideNJP}: each of the components of Eq. (\ref{hodgedecomp}) must thus verify independently this relation. 
This being trivially true for the curl and harmonic parts, it brings for the gradient term:
\begin{equation}
- \frac{1}{L_x} \frac{\partial^2 \varphi_\alpha}{\partial x^2}- \frac{1}{L_z} \frac{\partial^2 \varphi_\alpha}{\partial z^2} - \frac{1}{L_z} \frac{\partial^2 \varphi_\alpha}{\partial t'^2} =0 , \label{gradEq}
\end{equation}
which actually leads to Eqs. (\ref{propag}) reported in the previous Section. 

Eq. (\ref{gradEq}) and Eqs. (\ref{DeltaEq}) are formally equivalent. Consider a generic function $\varphi(x,z,t)$ veryfing this rule, and complying with periodic boundary conditions in all $x,z,t$. 
For symmetry reasons, if such a function would be used to describe the effect of a termination, we would impose a $k_{cx}$ sine or cosine for the transverse component, and a $\omega_\beta$ dependence for $t$.
Decomposing in a Fourier series the $z$-component, one finds then that {\it only a sine or cosine with wavevector} $\pm \beta$ is acceptable: by no means can we describe a decaying function within the guide.  
We therefore conclude that whatever perturbation is created in the charge-current vector by an end load attached to the guide, {\it it must be described by a curl term solely}.  
The specific writing of the Hodge decomposition for 
  TE$_{n=0,m}$ waves is discussed in Appendix \ref{topomodes};  topology and geometry are finally brought up in Section \ref{topodigr}. 

The adjoint and orthogonality properties are matched under proper 
{\it boundary conditions} only. 
For the scalar fields, the situation is trivial: all $\varphi_\alpha$ functions concerned by Eq. (\ref{varphialpha}) are periodic in $x,z,t$, as well as any of their derivatives. This means that we can {\it arbitrarily impose} periodic boundary conditions on our domain $[0,T]\times[-\ell,0]\times[-w,w]$. In turn, it guarantees a zero surface integral in the inner product Eq. (\ref{scal1}) when developing it with an integration by parts, which is the basis of the adjoint demonstration \cite{frankel}.
For the vector fields, it is more subtle. the gradient contributions are also periodic by construction and pose no problem. However, the $\vec{\Psi}_\alpha$ vectors must be chosen so as to guarantee zero boundary contributions as well. \\

We must thus proceed with an {\it Ansatz} for the allowed curl vectors. For a given mode indexed by $\alpha$, the curl components must obviously have the same time-dependence at frequency $\omega_\beta$, and transverse symmetry described by $k_{cx}$.  
We also impose the factorization of transverse $x$ and longitudinal $z$ dependencies, similarly to $\varphi_\alpha$. 
Likewise for symmetry reasons, the symmetric and antisymmetric bosonic operators must be equivalent: so they should share the same $z$-dependence in the $\vec{\Psi}_\alpha$ field.  This brings us to write, in the most generic fashion:
\begin{eqnarray}
\Psi_{x} & = & \phi_{zpf} \,  e^{+\mathbbm{i}\omega_\beta t}  \, \gamma_x(z) \times \label{Psix} \\
&& \!\!\!\!\!\!\!\!\!\!\!\!\!\!\!\!\!\!  \left[c^\dag_{s,\beta} \left( A_{\alpha,s,x} \cos[k_{cx}(x+\frac{w}{2})] + B_{\alpha,s,x} \sin[k_{cx}(x+\frac{w}{2})] \right) \right. \nonumber \\
&&  \!\!\!\!\!\!\!\!\!\!\!\!\!\!\!\!\!\!  \left. \!\!\! +c^\dag_{a,\beta} \left( A_{\alpha,a,x} \cos[k_{cx}(x+\frac{w}{2})] + B_{\alpha,a,x} \sin[k_{cx}(x+\frac{w}{2})] \right)  \right] \! , \nonumber \\
\Psi_{z} & = & \phi_{zpf} \,  e^{+\mathbbm{i}\omega_\beta t}  \, \gamma_z(z) \times \label{Psiz} \\
&& \!\!\!\!\!\!\!\!\!\!\!\!\!\!\!\!\!\!  \left[c^\dag_{s,\beta} \left( A_{\alpha,s,z} \cos[k_{cx}(x+\frac{w}{2})] + B_{\alpha,s,z} \sin[k_{cx}(x+\frac{w}{2})] \right) \right. \nonumber \\
&&  \!\!\!\!\!\!\!\!\!\!\!\!\!\!\!\!\!\!  \left. \!\!\! +c^\dag_{a,\beta} \left( A_{\alpha,a,z} \cos[k_{cx}(x+\frac{w}{2})] + B_{\alpha,a,z} \sin[k_{cx}(x+\frac{w}{2})] \right)  \right] \! , \nonumber \\
\Psi_{t} & = & \phi_{zpf} \,  e^{+\mathbbm{i}\omega_\beta t}  \, \gamma_t(z) \times  \label{Psit}\\
&& \!\!\!\!\!\!\!\!\!\!\!\!\!\!\!\!\!\!  \left[c^\dag_{s,\beta} \left( A_{\alpha,s,t} \cos[k_{cx}(x+\frac{w}{2})] + B_{\alpha,s,t} \sin[k_{cx}(x+\frac{w}{2})] \right) \right. \nonumber \\
&&  \!\!\!\!\!\!\!\!\!\!\!\!\!\!\!\!\!\!  \left. \!\!\! +c^\dag_{a,\beta} \left( A_{\alpha,a,t} \cos[k_{cx}(x+\frac{w}{2})] + B_{\alpha,a,t} \sin[k_{cx}(x+\frac{w}{2})] \right)  \right] \! , \nonumber  
\end{eqnarray} 
the proper physical current corresponding to the curl contribution is then obtained via $\{j_{curl,x},j_{curl,z},j_{curl,t}/(\mathbbm{i} \bar{v}) \}$  $ + c.c$.
We further simplify the problem by assuming that the $x-z$ factorization holds for the $\vec{j}_{curl}$ charge-current vector itself, such that this basic property is shared by {\it all the fields addressed} by the modeling. This is actually a rather strong assumption, which imposes $\gamma_x(z) = 
\gamma_z(z)=\gamma_t(z)=\gamma(z)$ and $d \gamma(z)/d z \propto \gamma(z)$.
Since the curl component describes the effect of the end load, {\it it must vanish inside the waveguide}. This leaves us with the solution:
\begin{eqnarray}
\gamma(z) &=& e^{+\kappa_\alpha z} \,\,\,\,\,\, \,\,\, \,\,\, \mbox{for $ z \approx 0$},\\
\gamma(z )&=&  e^{-\bar{\kappa}_\alpha (z+\ell)}\,\,\, \mbox{for $  z \approx  -\ell $} , 
\end{eqnarray}
reminding that formally, we shall require $\ell \gg 1/\kappa_{\alpha},1/\bar{\kappa}_\alpha$.
It also means that the two ends of the 
line are independent, and at $z \approx -\ell$ the constants appearing in Eqs. (\ref{Psix}-\ref{Psit}) should read $\bar{A}_{\alpha,i,j}, \bar{B}_{\alpha,i,j}$: the two connecting ends are presumably {\it different}.

We can now come back to the boundary condition issue.
It turns out that our $\vec{\Psi}_\alpha$ ansatz and its various derivatives are periodic in $x$ and $t$, but {\it not} in $z$: especially, $d \gamma(-\ell)/dz=-d \gamma(0)/dz$. 
One neat  way of fixing this problem is to extend our $z$ domain to $+\ell/2$ on the positive side and  $-3\ell/2$ on the negative side, and ask then the $\gamma(z)$ function to be continued there in a $C^\infty$ fashion such that $\gamma(\ell/2)=\gamma(-3\ell/2)=1+\epsilon$ (with say $0<\epsilon<1$), and $d^n \gamma(\ell/2)/dz^n=d^n \gamma(-3\ell/2)/dz^n=0$ for $n>0$. Such a function can then be replicated with a $2\ell$ periodicity, 
which makes it as periodic {\it and} regular as the sine and cosine functions on the new domain (except for the irrelevant discontinuity at $z=-\ell/2$; see Fig. \ref{fig4} in Appendix \ref{topomodes} for an illustration). It suffice then to extend the $z-$integral in the scalar product Eq. (\ref{scal2}) to the new boundaries, and impose again periodic boundary conditions on $z$ for the space of vector functions considered. 
Hodge's adjoint properties are then guaranteed, with the produced fields physically meaningful only within $-\ell<z<0$ (similarly to the argument invoked above for $g_\phi$ on the $[-w,w]$ domain). \\

The sum $\vec{j}=\sum_\alpha \vec{j}_\alpha$ produces then the 
sought solution.
Essentially, at that stage the only assumption made is the introduction of {\it a single decay rate} $\kappa_\alpha, \bar{\kappa}_\alpha$ characterizing each end of the line.
These are outside of our modeling, and arise from the complex electromagnetic field structure confined in the waveguide at the level of the connecting ends. Finite element simulations would be required to actually describe these zones, and verify how good our approximation might be in a given specific case.
Let us for now consider only the $z=0$ interface, the reasoning being strictly equivalent on the other one.
 For us, the connector's healing length $1/\kappa_\alpha$ is a given. 
Note that the $\vec{\Psi}_\alpha$ vectors are not unique, and we are left with the task of finding proper $A_{\alpha,i,j}, B_{\alpha,i,j}$ (complex) coefficients matching the boundary conditions at $z=0$.
The questions of {\it existence and unicity} of our solution, which are usually guaranteed by  proper scalar products, are discussed in Section \ref{topodigr}.

\section{Load description}
\label{load}

Following the same logic as for the curl term in the previous Section, we must decompose the charges and currents at the level of the termination $z=0$ into a series summing over $\alpha$, namely $\vec{j}(x,z=0,t)=\sum_\alpha \vec{j}_{load}(x,t)$. TE$_{n=0,m}$ waves are again excluded from the sum and treated specifically in Appendix \ref{topomodes}.
The generic writing of these components can be produced as: 
\begin{eqnarray}
j_{load,x}(x,t) & = & \mathbbm{i} k_{cx}\, \frac{\phi_{zpf} \,  e^{+\mathbbm{i}\omega_\beta t} }{L_x} \times \nonumber  \\
&& \left[ \tilde{g}_\phi(x) \, {\cal C}_{\alpha,x}\, d_{c,x}^\dag +g_\phi(x) \, {\cal D}_{\alpha,x} \, d_{d,x}^\dag \right] \nonumber \\
&& + c.c. , \label{loadx} \\
j_{load,z}(x,t) & = & \beta\, \frac{\phi_{zpf} \,  e^{+\mathbbm{i}\omega_\beta t} }{L_z} \times \nonumber \\
&& \left[ g_\phi(x) \, {\cal C}_{\alpha,z} \, d_{c,z}^\dag + \tilde{g}_\phi(x) \, {\cal D}_{\alpha,z} \, d_{d,z}^\dag \right] \nonumber \\
&& + c.c. ,  
\end{eqnarray}
\begin{eqnarray}
\sigma_{load}(x,t) & = & \omega_\beta C_z \, \phi_{zpf} \,  e^{+\mathbbm{i}\omega_\beta t} \times \nonumber \\
&& \left[g_\phi(x) \, {\cal C}_{\alpha,t} \, d_{c,t}^\dag+ \tilde{g}_\phi(x) \, {\cal D}_{\alpha,t} \, d_{d,t}^\dag \right] \nonumber \\
&& + c.c. \label{loads} ,
\end{eqnarray}
with $j_{load,t}=\mathbbm{i} \bar{v} \, \sigma_{load}$ and the function $\tilde{g}_\phi$ such that:
\begin{eqnarray}
\tilde{g}_\phi(x) & = & \cos \left[ k_{cx} \left(x+\frac{w}{2}\right)\right] \,\, \mbox{for $n>0$}, \\
\tilde{g}_\phi(x) & = & 0 \,\,\,\, \mbox{for $n=0$}. 
\end{eqnarray}
In the above, the dimensionless (complex) constants ${\cal C}_{\alpha,i}, {\cal D}_{\alpha,i}$ characterize the load, and the operators $d_i^\dag, d_i$ 
must be localized  bosonic creation/annihilation operators.
At that stage, it is already obvious that many of these are in surplus: {\it only two different} output bosons are required in order to write a scattering matrix.

The charge-current vector $\vec{j}_{\alpha}$ can be recast at $z=0$ into:
\begin{eqnarray}
j_{\alpha,x}(x,z=0,t) & = & \mathbbm{i} k_{cx}\, \frac{\phi_{zpf} \,  e^{+\mathbbm{i}\omega_\beta t} }{L_x} \times \nonumber  \\
&& \!\!\!\!\!\!\!\!\!\!\!\! \left( \tilde{g}_\phi(x) \left[ \left(\sqrt{2}+ {\cal K}_{\alpha,s,x}\right)\, c_{s,\beta}^\dag + {\cal K}_{\alpha,a,x}\, c_{a,\beta}^\dag  \right] \right. \nonumber \\
&& \left. +g_\phi(x) \left[  {\cal L}_{\alpha,s,x} \, c_{s,\beta}^\dag + {\cal L}_{\alpha,a,x}\, c_{a,\beta}^\dag  \right] \right) \nonumber \\
&& + c.c. , \label{jax} \\
j_{\alpha,z}(x,z=0,t) & = & \beta\, \frac{\phi_{zpf} \,  e^{+\mathbbm{i}\omega_\beta t} }{L_z} \times \nonumber \\
&&\!\!\!\!\!\!\!\!\!\!\!\!  \left( g_\phi(x) \left[ {\cal K}_{\alpha,s,z}\, c_{s,\beta}^\dag + \left(\sqrt{2}+ {\cal K}_{\alpha,a,z}\right)\, c_{a,\beta}^\dag \right] \right. \nonumber \\
&& \left. +\tilde{g}_\phi(x) \left[{\cal L}_{\alpha,s,z} \, c_{s,\beta}^\dag + {\cal L}_{\alpha,a,z}\, c_{a,\beta}^\dag  \right] \right) \nonumber \\
&& + c.c. , \\
\sigma_{\alpha}(x,z=0,t) & = &  \omega_\beta C_z \, \phi_{zpf} \,  e^{+\mathbbm{i}\omega_\beta t} \times \nonumber \\
&& \!\!\!\!\!\!\!\!\!\!\!\!  \left( g_\phi(x) \left[ \left(\sqrt{2}+{\cal K}_{\alpha,s,t}\right)\, c_{s,\beta}^\dag +  {\cal K}_{\alpha,a,t} \, c_{a,\beta}^\dag \right] \right. \nonumber \\
&& \left. +\tilde{g}_\phi(x) \left[ {\cal L}_{\alpha,s,t} \, c_{s,\beta}^\dag + {\cal L}_{\alpha,a,t}\, c_{a,\beta}^\dag  \right] \right) \nonumber \\
&& + c.c. \label{jas} ,
\end{eqnarray}
with $j_{\alpha,t}=\mathbbm{i} \bar{v} \, \sigma_{\alpha} $ and the coefficients ${\cal K}_{\alpha,i,j},{\cal L}_{\alpha,i,j}$ being deduced from the $A_{\alpha,i,j},B_{\alpha,i,j}$ of the curl expression.
Note that for TEM and TM$_{n=0,m}$, the problem at hand simplifies notably: there is no $x$ component above, and $\tilde{g}_\phi=0$ in the $z,t$ expressions.
But for a common TE$_{n\neq0,m}$ branch, {\it these generic formulas must be considered}.

The compatibility between Eqs. (\ref{loadx}-\ref{loads}) and Eqs. (\ref{jax}-\ref{jas}) leads to a set of equations, when projected onto the $g_\phi, \tilde{g}_\phi$ functions. 
The surplus equations involving ${\cal L}_{\alpha,i,j}, {\cal D}_{\alpha,i}$ cannot be matched concurrently with the other ones, so we are forced to fix:
\begin{equation}
{\cal L}_{\alpha,i,j} = {\cal D}_{\alpha,i} =0. \label{impose1}
\end{equation}
This already removes some of our supernumerary localized bosons, and simply means that the load boundary condition for given $\alpha$ {\it must match the transverse wave pattern of the transmitted signals}.
We are then left with:
\begin{eqnarray}
  \left(\sqrt{2}+ {\cal K}_{\alpha,s,x}\right)\, c_{s,\beta}^\dag + {\cal K}_{\alpha,a,x}\, c_{a,\beta}^\dag    & = &  {\cal C}_{\alpha,x}\, d_{c,x}^\dag , \label{eq1} \\
  {\cal K}_{\alpha,s,z}\, c_{s,\beta}^\dag + \left(\sqrt{2}+ {\cal K}_{\alpha,a,z}\right)\, c_{a,\beta}^\dag   & = &  {\cal C}_{\alpha,z}\, d_{c,z}^\dag , \label{eq2} \\
  \left(\sqrt{2}+{\cal K}_{\alpha,s,t}\right)\, c_{s,\beta}^\dag +  {\cal K}_{\alpha,a,t} \, c_{a,\beta}^\dag   &= &  {\cal C}_{\alpha,t}\, d_{c,t}^\dag ,\label{eq3}
\end{eqnarray}
and the complex conjugate ones.
In order to map this problem onto a scattering matrix, we must impose that Eq. (\ref{eq1}) {\it is equivalent} to Eq. (\ref{eq3}): the transverse current generated by TE modes does not carry independent quantum information. This leads to:
\begin{eqnarray}
 {\cal K}_{\alpha,s,x} & = & {\cal K}_{\alpha,s,t} , \label{impose2}\\
 {\cal K}_{\alpha,a,x} & = & {\cal K}_{\alpha,a,t}, \label{impose3}\\
 {\cal C}_{\alpha,x} & = &   {\cal C}_{\alpha,t}  , \label{CxCt}
\end{eqnarray}
and obviously:
\begin{equation}
d_{c,x}^\dag = d_{c,t}^\dag \,\, , \,\, d_{c,x}  = d_{c,t} .
\end{equation}
We are now left with two independent equations, and only two independent localized bosons ($d_{c,z}^\dag,d_{c,z}$ and $d_{c,t}^\dag,d_{c,t}$) coupled to the waveguide ones ($c_{s,\beta}^\dag,c_{s,\beta}$ and $c_{a,\beta}^\dag,c_{a,\beta}$). As is required.

Let us finally introduce from Eqs. (\ref{sigma},\ref{jz}) the localized charge per unit length  and current at the boundary:
\begin{eqnarray}
q_{load} (t) & = & \omega_\beta C_z \, \phi_{zpf} \,  e^{+\mathbbm{i}\omega_\beta t} w_{ef\!f} \, {\cal C}_{\alpha,t} \,  d_{c,t}^\dag + c.c. , \\
I_{load} (t) & = & \beta\, \frac{\phi_{zpf} \,  e^{+\mathbbm{i}\omega_\beta t} }{L_z} w_{ef\!f} \, {\cal C}_{\alpha,z} \,  d_{c,z}^\dag + c.c. 
\end{eqnarray}
Eq. (\ref{U}) enables then to define the load voltage from the relationship $q= w_{ef\!f} C_z \, U$:
\begin{equation}
U_{load}(t) = \omega_\beta   \, \phi_{zpf} \,  e^{+\mathbbm{i}\omega_\beta t}   \, {\cal C}_{\alpha,t} \,  d_{c,t}^\dag + c.c.
\end{equation}
In order to restore a more classical writing, we can formally re-express these formulas in terms of {\it complex amplitudes}:
\begin{eqnarray}
U_{load} (t) & = & U_0 \, e^{+\mathbbm{i}\omega_\beta t} \,  d_{c,t}^\dag + U_0^*\, e^{-\mathbbm{i}\omega_\beta t}\,  d_{c,t} , \\
I_{load} (t) & = & I_0 \, e^{+\mathbbm{i}\omega_\beta t} \,  d_{c,z}^\dag + I_0^* \, e^{-\mathbbm{i}\omega_\beta t} \,  d_{c,z} ,
\end{eqnarray}
which brings in the conventional definition of a {\it load impedance}:
\begin{equation}
Z_{load} = \frac{U_0}{I_0} = Z_0 \, \frac{{\cal C}_{\alpha,t}}{{\cal C}_{\alpha,z}} , \label{loadZ}
\end{equation}
with $Z_0$ the characteristic impedance Eq. (\ref{caracZ}).
Injecting Eqs. (\ref{impose1}, \ref{CxCt},\ref{loadZ}) into Eqs. (\ref{loadx}-\ref{loads}) we realize that  
each $\alpha$ component of the load is solely given by a complex number $ {\cal C}_{\alpha,z} $ and the ratio $Z_{load}/Z_0$ (also complex). 
These are characteristic of the mode $\alpha$; but of course, different modes can share the same values (as is the case in the conventional TEM model).
Note that one could have equivalently introduced a load admittance $Y_{load} = 1/Z_{load}$, a characteristic admittance $Y_0=1/Z_0$, and rewritten the load charge-current $\vec{j}_{load}$ in terms of $Y_{load}/Y_0$ and $ {\cal C}_{\alpha,t} $. \\

We must now make the link with the scattering matrix formalism, and see what it implies for  $ {\cal C}_{\alpha,z} $ and the  $ {\cal K}_{\alpha,i,j} $ constants. Obtaining a meaningful theory that reproduces {\it both} the results of classical engineering and of quantum scattering is the aim of the next Section.

\section{Solving for a scattering matrix}
\label{scatter}

Our $c_{s,\beta}^\dag,c_{s,\beta}$ and $c_{a,\beta}^\dag,c_{a,\beta}$ represent specifically symmetric and anti-symmetric standing wave modes constructed from the traveling bosons $b_{\beta}^\dag,b_{\beta}$ and $b_{-\beta}^\dag,b_{-\beta}$. Similarly, our load operators must originate from traveling ones:
\begin{equation}
\!\!\!\!\!\!\!\!\!\!  
\begin{pmatrix}
d_{c,z}^\dag \\
d_{c,t}^\dag 
\end{pmatrix}  
= e^{+ \mathbbm{i} \varphi_0} \!\!
\begin{pmatrix}
-e^{+ \mathbbm{i} \varphi_1} \sin(\Theta) & e^{+ \mathbbm{i} \varphi_2}\cos(\Theta) \\
e^{- \mathbbm{i} \varphi_2}\cos(\Theta) & e^{- \mathbbm{i} \varphi_1}\sin(\Theta)
\end{pmatrix} \!\!
\begin{pmatrix}
a_{in}^\dag \\
a_{out}^\dag 
\end{pmatrix} \! ,
\end{equation}
together with the equivalent conjugate relationship.
But for the time being, {\it no hypothesis is made on the symmetries} of this combination, and the
 transformation is kept universal (while preserving bosonic properties).
 $a_{in}^\dag,a_{in}$ and $a_{out}^\dag,a_{out}$ represent incoming and outgoing bosons from another guide connected at $z=0$, or bath modes from a physical dissipative load (Caldeira and Leggett \cite{leggett}) maintained at a given temperature (Fig. \ref{fig1}).
In order to match conventional terminology, we  rewrite $b_\beta^\dag = b_{in}^\dag$ and $b_{-\beta}^\dag = b_{out}^\dag$ (and similarly without the dagger).
We can then recast Eqs. (\ref{eq2},\ref{eq3}) into:
\begin{eqnarray}
& & b_{in}^\dag \left( 1+ \bar{{\cal K}}_{\alpha,z} \right) + b_{out}^\dag \left(- 1+ \Delta{\cal K}_{\alpha,z} \right)=  \label{aout} \\
& & \!\!\!\!\!\!\!  {\cal C}_{\alpha,z} e^{+ \mathbbm{i} \varphi_0} \! \left[ -e^{+ \mathbbm{i} \varphi_1} \sin(\Theta)\, a_{in}^\dag +e^{+ \mathbbm{i} \varphi_2} \cos(\Theta)\,a_{out}^\dag  \right] \! ,\nonumber  \\
& & b_{in}^\dag \left( 1+ \bar{{\cal K}}_{\alpha,t} \right) + b_{out}^\dag \left( +1+ \Delta{\cal K}_{\alpha,t} \right) =  \label{bout} \\
& &  \!\!\!\!\!\!\!  {\cal C}_{\alpha,z} e^{+ \mathbbm{i} \varphi_0} \frac{Z_{load}}{Z_0} \left[e^{- \mathbbm{i} \varphi_2} \cos(\Theta)\, a_{in}^\dag +e^{- \mathbbm{i} \varphi_1} \sin(\Theta)\,a_{out}^\dag \right]\! ,\nonumber
\end{eqnarray}
with:
\begin{eqnarray}
\bar{{\cal K}}_{\alpha,z}& =& \frac{{\cal K}_{\alpha,s,z} + {\cal K}_{\alpha,a,z}}{\sqrt{2}} , \label{kacal1} \\
\Delta{\cal K}_{\alpha,z}& =& \frac{{\cal K}_{\alpha,s,z} - {\cal K}_{\alpha,a,z}}{\sqrt{2}} , \\
\bar{{\cal K}}_{\alpha,t}& =& \frac{{\cal K}_{\alpha,s,t} + {\cal K}_{\alpha,a,t}}{\sqrt{2}} , \\
\Delta{\cal K}_{\alpha,t}& =& \frac{{\cal K}_{\alpha,s,t} - {\cal K}_{\alpha,a,t}}{\sqrt{2}} . 
\end{eqnarray}
The phase $\varphi_0$ can thus be incorporated in the defintion of $ {\cal C}_{\alpha,z}$ with no loss of generality.  

Eqs. (\ref{impose1},\ref{impose2},\ref{impose3}) bring constraints on the curl coefficients $A_{\alpha,i,j}, B_{\alpha,i,j}$. In particular, it turns out that:
\begin{eqnarray}
\bar{{\cal K}}_{\alpha,t}& =& \mathbbm{i} \eta \, \bar{{\cal K}}_{\alpha,z} , \\
\Delta{\cal K}_{\alpha,t}& =& \mathbbm{i} \eta \, \Delta{\cal K}_{\alpha,z} , \label{kacal2}
\end{eqnarray}
with $\eta = \kappa_\alpha/\beta$ in {\it all} configurations (a given positive number in the model). 
Only two unknown complex coefficients are thus left free in the curl expansion.
Eqs. (\ref{aout},\ref{bout}) can be reshaped in the standard matrix form:
\begin{equation}
\begin{pmatrix}
a_{out}^\dag \\
b_{out}^\dag 
\end{pmatrix}  
= 
\begin{pmatrix}
\mathbf{r}' & \mathbf{t} \\
\mathbf{t}' & \mathbf{r} 
\end{pmatrix} \!\!
\begin{pmatrix}
a_{in}^\dag \\
b_{in}^\dag 
\end{pmatrix} \! , \label{scattering}
\end{equation}
which must be turned into a scattering matrix.
The parameters introduced here, together with the curl $A_{\alpha,i,j},B_{\alpha,i,j}$ ones, are listed in Appendix \ref{curlparams} for the interested reader.

Eq. (\ref{scattering}) is compliant with our requirements if, and only if:
\begin{eqnarray}
|\mathbf{r}|^2 +|\mathbf{t}|^2 &= & 1 , \\
|\mathbf{r}'|^2 +|\mathbf{t}'|^2 & = & 1 , \\
\mathbf{t} \, \mathbf{r}'^* + \mathbf{r} \, \mathbf{t}'^* & = & 0 , \\
\mathbf{r} & = & \frac{Z_{load}-Z_0}{Z_{load}+Z_0} .
\end{eqnarray}
The three first conditions mean that the process is unitary (i.e. lossless).
Note that $\mathbf{t}=\mathbf{t}'$ is {\it not} imposed here, meaning that the 2 port junction can be non-reciprocal ($|t|=|t'|$ but a {\it phase} difference can exist between $\mathbf{t}$ and $\mathbf{t}'$). 
The last equation above implements the classical result.

Let us pose, for convenience:
\begin{eqnarray}
\frac{Z_{load}}{Z_0} & = & r_z \, e^{+ \mathbbm{i} \varphi_z} , \\
{\cal C}_{\alpha,z} & = & {\cal C}_0 \, e^{+ \mathbbm{i} \varphi_0} , \\
\varphi_1+\varphi_2 & = & \varphi_z + \epsilon , \\
\varphi_{-} & = & \varphi_0 - \varphi_1 , \\
\varphi_{+} & = & \varphi_0 + \varphi_1 ,
\end{eqnarray}
with $r_z, {\cal C}_0 >0$, and $r_z, \varphi_z$ given {\it for the specific mode} $\alpha$ under study.
It turns out that it is {\it always} possible to find $\Theta(\epsilon)$ that satisfies our requirements. 
Interestingly:
\begin{eqnarray}
\epsilon & = & 0 , \\
\Theta & = & \frac{\pi}{4} , 
\end{eqnarray}
is a valid solution {\it for any} $r_z, \varphi_z$. 
One can thus chose this option with no loss of generality. The scattering matrix coefficients then write:
\begin{eqnarray}
\mathbf{r} &=& 1-\frac{2}{1+r_z \, e^{ \mathbbm{i} \varphi_z}} , \label{solr} \\
\mathbf{t} &=& 2 \sqrt{2} \,\frac{e^{- \mathbbm{i} \left( \varphi_- + \varphi_z\right)}}{{\cal C}_0} \frac{\left( r_z \, e^{ \mathbbm{i} \varphi_z} - \mathbbm{i} \eta \right)}{\left( r_z \, e^{ \mathbbm{i} \varphi_z}+1\right)\left( r_z- \mathbbm{i} \eta\right)} , \\
\mathbf{r}' &=& e^{- \mathbbm{i} \left( \varphi_- - \varphi_+  + \varphi_z\right)} \times \nonumber \\
&& \frac{\left(  e^{ \mathbbm{i} \varphi_z} - r_z \right)\left(r_z \,e^{ \mathbbm{i} \varphi_z} - \mathbbm{i} \eta  \right)\left( \eta - \mathbbm{i}  r_z \right)}{\left( r_z \, e^{ \mathbbm{i} \varphi_z}+1\right)\left( r_z- \mathbbm{i} \eta\right) \left(\eta \, e^{ \mathbbm{i} \varphi_z} - \mathbbm{i} r_z \right) } , \\
\mathbf{t}' & = & \frac{{\cal C}_0 \, e^{ \mathbbm{i} \varphi_+} }{\sqrt{2} } \frac{r_z \left(1+ e^{ 2\mathbbm{i}  \varphi_z} \right) \left( \eta-\mathbbm{i} r_z\right) }{\left( r_z \,e^{ \mathbbm{i}  \varphi_z} + 1 \right)\left(\eta \, e^{ \mathbbm{i} \varphi_z} - \mathbbm{i} r_z \right)} , \label{soltp}
\end{eqnarray}
with:
\begin{equation}
{\cal C}_0  =  \sqrt{\frac{2}{ r_z \, \cos(\varphi_z)}} \sqrt{1-\frac{2 r_z \, \eta \sin(\varphi_z)}{r_z^2+ \eta^2 }} . \label{Eqc0}
\end{equation}
In Eqs. (\ref{solr}-\ref{soltp}), the two phases $\varphi_- , \varphi_+$ are left free and characterize the load bosons; only for a reciprocal configuration shall a single phase variable remain.   
From Eq. (\ref{Eqc0}), we see that we must impose $\cos(\varphi_z) > 0$, which simply means that the real part of our load {\it  should be strictly positive}.
All required constants are thus fixed, and 
the found solution for $\bar{{\cal K}}_{\alpha,z},\Delta{\cal K}_{\alpha,z}$ is given in Appendix \ref{curlparams}. 
Note that these parameters are {\it nonzero} functions of $\eta,r_z$ and $\varphi_z$; the values vanish only for $\varphi_z=0$.  \\

Let us end the Section by commenting on the usual limits. One easily verifies that when $r_z \rightarrow 0^+$, one gets $\mathbf{r}=-1, \mathbf{r}' = +e^{- \mathbbm{i} \left( \varphi_- - \varphi_+  + \varphi_z\right)}$ and $\mathbf{t}=\mathbf{t}'=0$.
Conversely, $r_z \rightarrow +\infty$ leads to $\mathbf{r}=+1, \mathbf{r}' = - e^{- \mathbbm{i} \left( \varphi_- - \varphi_+  + \varphi_z\right)}$ and also 
$\mathbf{t}=\mathbf{t}'=0$.
This means $a_{out}^\dag \equiv a_{in}^\dag$ and $b_{out}^\dag \equiv b_{in}^\dag$ up to a phase factor (meant by the $\equiv$ symbol, and similarly for conjugates).
These are our conventional "short" or "open" configurations respectively, which can be mapped one onto the other by swapping the description from impedances $Z_i$ to admittances $Y_i$.
Taking $r_z=1$ and $\varphi_z=0$, one obtains the usual "matched load", with $\mathbf{r}= \mathbf{r}' =0$ and $\mathbf{t}=+e^{- \mathbbm{i}  \varphi_- }, \mathbf{t}' =+e^{+ \mathbbm{i}  \varphi_+ } $. Then $a_{out}^\dag \equiv b_{in}^\dag$ and $b_{out}^\dag \equiv a_{in}^\dag$ (up to a phase factor).

\section{A digression on topology, geometry and gauge}
\label{topodigr}

Hodge theory elegantly relates analysis, geometry and topology. 
The formulation presented here reproduces the ingredients that it requires, apart from an important one: we {\it do not make use of a proper scalar product}. So, what do we miss? 
Essentially the demonstration of {\it existence and unicity} of the Hodge decomposition {\it for any} charge-current vector.
But in practice, we {\it did present} such a solution to our specific problem: so it exists. Besides, the whole demonstration performed has been realized with {\it only one} genuine assumption: a single decay rate $\kappa_\alpha$ (per mode) for the loads' impact. Therefore, the found solution is, in this respect, unique. So to speak, we do not require existence nor unicity for any other field. 

The harmonic term has been kept outside of our discussion; it does not play any role in the load problem (and {\it topology} is not an ingredient of our theory). But thanks to Hodge's decomposition,  
a clear {\it geometrical} difference emerges between TEM, TM, TE$_{n\neq0,m}$ and  TE$_{n=0,m}$ charge-currents transporting quantum information: 
the former are {\it gradient-like}, while the latter are {\it  curl-like} (see Appendix \ref{topomodes}). 
This certainly has to be related to the fields' specific gauge attributes. 
Indeed, the Devoret relationship $\Delta V = \partial \varphi/\partial t$ presumes a {\it fixed gauge} in which $\Delta V$ is meaningful \cite{EddyguideNJP,AlexguideNJP}. 
For all conventional waves, this is no problem: they all verify an internal gauge symmetry where the so-called transverse and longitudinal gauges can play interchangeable roles for $\varphi$ (a gauge transformation swapping from one to the other); the Devoret result being a consequence of the longitudinal one. However, TE$_{n=0,m}$ modes {\it break this symmetry} \cite{AlexguideNJP}. In the Cartesian geometry, it happens to be nonetheless possible to produce a proper longitudinal gauge; but this is {\it impossible} for cylindrical waveguides. 
One can still apply an "effective" Devoret expression, simply by {\it defining} the voltage $U$ through the charge with $q= w_{ef\!f} C_z \, U$.
This conveniently preserves all our conventional framework; but {\it by no means} does it carry any gauge information.
What this internal gauge symmetry actually means remains an open question.
Similarly, {\it virtual electrodes} had to be invoked in order to formalize the TE$_{n=0,m}$ situation. But in cylindrical geometries, the de Rham complex is defined only on the sub-space of {\it solutions to our problem}, and not everywhere (Appendix \ref{topomodes}). Nonetheless, a "Hodge" decomposition can still be produced, and seems unique. What this other peculiarity of  TE$_{n=0,m}$ waves means shall also be left for future work.

\section{Conclusion}
\label{conclu}

The model presented in this manuscript enables to map the engineering problem of microwaves traveling in a waveguide \cite{pozar} onto a quantum scattering matrix one \cite{devoretRMP}. The procedure applies equally well to {\it all types} of waves propagating in square-type and cylindrical geometries: the conventional TEM, but also TM and TE modes. This is performed through {\it the quantization of a generalized flux field} defined on the electrodes \cite{devoret}, following our previous works \cite{EddyguideNJP,AlexguideNJP}.
It is based on the use of Hodge's decomposition theorem \cite{frankel}, that enables to separate the charge-current vector flowing in the metallic boundaries confining the fields into {\it gradient-like} and {\it curl-like} terms.

It turns out that TEM, TM and TE$_{n>0,m}$ families are all described by gradient charge-current traveling fields. On the other hand, TE$_{n=0,m}$ modes are  curl. This striking difference should be set against another one: the {\it gauge symmetry} that is broken by TE$_{n=0,m}$ waves \cite{AlexguideNJP}.
The fundamental meaning of this symmetry, and its potential use for quantum information transfer, remains an open question.

The perturbation due to the termination is itself described by a curl charge-current field. It is characterized here by {\it one healing length} $1/\kappa_{\alpha}$ per mode $\alpha$ that quantifies how much this perturbation "leaks" into the guide. It leads to a simple decomposition of what a {\it generic load} must be: a series summing over all modes, and sharing each mode's transverse pattern (of wavevector $k_{cx}$), creating the voltage $U(x,z=0,t)$ and current $I(x,z=0,t)$ localized at the end of the line.
Within this series, a {\it single complex coefficient} $Z_{load}(\alpha)$ characterizes each mode pattern, with 
$Re(Z_{load})>0$. Two phases $\varphi_+,\varphi_-$ are constitutive of the output bosons, fixing the most generic scattering matrix; even if non-reciprocal.  

The values of $\kappa_{\alpha}$, $Z_{load}(\alpha)$ and 
$\varphi_+,\varphi_-$ are outside of the model, and need to be obtained from a specific experimental configuration. 
We do believe that most of the ideas developed here are quite generic; future works shall extend the theory to other types of waveguides, like e.g. coplanar ones.
Furthermore, considering more than 2 ports joining together (like e.g. in a circulator) would be very useful. 
Finally, the whole modeling treats time $t$ as a global variable: making the connection with {\it quantum pulses} shall certainly be fruitful \cite{molmer,quantumpulses}.

\vspace*{1cm}
(\dag) Corresponding Author: eddy.collin@neel.cnrs.fr

\begin{acknowledgements}

The Authors acknowledge valuable support from the N\'eel laboratory and the 
 European Microkelvin Platform (EMP) consortium, 
visit: \underline{https://emplatform.eu/}.

\end{acknowledgements}

\section{Data availability}

No data was used or created for this manuscript. A Mathematica\textsuperscript{\textregistered}$\,$ code is available at the following URL: 
\small{\underline{https://cloud.neel.cnrs.fr/index.php/s/CnnYPKn8XHYZgXa}}.

\appendix

\section{Case of the TE$_{n=0,m}$ modes}
\label{topomodes} 

The charge-current vector of TE$_{n=0,m}$ modes (denoted TE$_0$ below) is radically different from the other ones. It can be shaped into \cite{EddyguideNJP,AlexguideNJP}:
\begin{eqnarray}
j_{T\!E_0,x} & = & -(-1)^q \, \frac{1}{L_x'} k_c \, \varphi_\alpha' (x,z,t) ,\\
j_{T\!E_0,z} & = & 0 ,\\
j_{T\!E_0,t} & = & 0 ,
\end{eqnarray}
with $q=1,2$ depending on the mode, and the function $\varphi_\alpha'$ defined from:
\begin{eqnarray}
\varphi_\alpha'(x,z,t) & = & \varphi_\beta' (x,z,t)+\varphi_{-\beta}' (x,z,t) , \\
\varphi_\beta' (x,z,t) & = & \phi_{zpf} \, g_\phi(x) \, \tilde{f}_\beta (z,t) , \label{phiprime} \\
  g_\phi(x) & = & 1 .
\end{eqnarray}
By construction, $k_{cx}=0$ and $k_c = k_{c\eta}$.
One must understand that at that stage, while tempting, this function {\it is not} of the same nature as the $\varphi_\beta$ one presented in Eq. (\ref{varphi1}). 
Indeed, this charge-current vector is not a gradient form, but rather a {\it curl}. Consider:
\begin{eqnarray}
\Psi_{x}' & = & 0, \\
\Psi_{z}' & = & -\frac{\partial \varphi_{rot} (x,z,t)}{\partial t'}, \\
\Psi_{t}' & = & +\frac{\partial \varphi_{rot} (x,z,t)}{\partial z},
\end{eqnarray}
with:
\begin{equation}
\varphi_{rot} (x,z,t) = -(-1)^q \,\frac{L_x}{L_x'\, k_c} \varphi_\alpha'(x,z,t) ,
\end{equation}
for normalization purposes. One easily shows that 
$\vec{j}_{T\!E_0}$ is the curl of $\vec{\Psi}'$
when using relations Eqs. (\ref{curl1}-\ref{curl3}).
Note that in this specific case, nothing is imposed to the inductance coefficients defined for our electrode, so we can safely impose $L_x=L_z=L_x'$ making the writing simpler. And $C_z$ (and thus $\bar{v}$) can be any.

Following the same reasoning as for the other types of waves, we pose the load current-charge vector as:
\begin{eqnarray}
j_{load,x}(x,t) & = & \mathbbm{i} k_{c}\, \frac{\phi_{zpf} \,  e^{+\mathbbm{i}\omega_\beta t} }{L_x'} \times \nonumber  \\
&& \left[ g_\phi(x) \, {\cal D}_{\alpha,x} \, d_{d,x}^\dag \right] \nonumber \\
&& + c.c. ,   \\
j_{load,z}(x,t) & = & 0  , \\
j_{load,t}(x,t) & = & 0 .
\end{eqnarray}
The perturbation due to the termination must again be described by a curl form implying a $\vec{\Psi}_{\alpha}$. The continuity at $z=0$ reads then:
\begin{equation}
   \left(\sqrt{2}+ {\cal K}_{\alpha,s,x}\right)\, c_{s,\beta}^\dag + {\cal K}_{\alpha,a,x}\, c_{a,\beta}^\dag      =  {\cal D}_{\alpha,x}\, d_{d,x}^\dag  . \label{redunt}
\end{equation}
 We obtain {\it a single} equation, which looks like Eq. (\ref{eq1}). 
This is obviously not enough to construct a scattering matrix, and similarly to the strategy applied to conventional waves, it must be redundant with a proper set of equations that can describe the load. The found coefficients are given in Appendix \ref{curlparams}.
 
This proper set is obtained thanks to the introduction of {\it virtual electrodes} \cite{EddyguideNJP,AlexguideNJP}.
The idea is that {\it we do have} planes in the guide's geometry that support standard metallic boundary conditions leading to longitudinal currents and surface charges. 
For the parallel plates, these planes are the ones closing the guide on its side. For a cylinder, these are diameter planes. Only for the rectangular guide, where these planes are simply the other set of electrodes (they close the guide, as for the parallel plate configuration) do real electrodes exist there. But for all other virtual cases, one should think of these planes as surfaces where one would detect currents and charges {\it if a non-invasive detector} was immersed in the field, at that specific place in space.
We can then define a "virtual" generalized flux field living on these planes:
\begin{eqnarray}
\varphi_{v,\alpha} (x,z,t) & = & \varphi_{v,\beta} (x,z,t)+\varphi_{v,-\beta} (x,z,t) , \\
\varphi_{v,\beta}(x,z,t) &  =  & \phi_{zpf} \, g_v(x) \, \tilde{f}_{\beta}(z,t) , \label{gtilde}
\end{eqnarray}
the difference with Eq. (\ref{varphi1}) being in the $g_v$ function. The charge-current associated vector reads:
\begin{eqnarray}
j_{v,x} (x,z,t') & = & -\frac{1}{L_x} k_c \,  \tilde{\varphi}_{v,\alpha} (x,z,t) , \label{pathos} \\
j_{v,z} (x,z,t') & = & -\frac{1}{L_z} \frac{\partial  \varphi_{v,\alpha} (x,z,t)}{\partial z }, \\
j_{v,t} (x,z,t') & = & -\frac{1}{L_z} \frac{\partial  \varphi_{v,\alpha} (x,z,t)}{\partial t' } ,
\end{eqnarray}
with $\tilde{\varphi}_{v,\alpha}$ written like Eq. (\ref{gtilde}), but exchanging $g_v$ for $\tilde{g}_v$. Note that the inductances $L_x=L_z$ (verified for TE$_0$ modes) and the corresponding capacitance $C_z$ are defined here for the virtual plane; the real electrode parameters discussed above will require to be properly linked to these (see below).

In the Cartesian geometries (parallel plates and rectangular guide), the situation is trivial. 
We simply have $k_{c\eta}=k_{cy}$ and:
\begin{eqnarray}
g_v(x) & = & \sin \left[ k_{cy} \left(x+\frac{d}{2}\right)\right]  , \\
\tilde{g}_v(x) & = & \cos \left[ k_{cy} \left(x+\frac{d}{2}\right)\right]  ,
\end{eqnarray}
with $k_{cy}= m \, \pi/d$ and $d$ the height of the guide (having always $m>0$).
This simply means that Eq. (\ref{pathos}) can be written in terms of a gradient $x$-component $\partial/\partial x$. The problem at hand is thus strictly equivalent to the one treated in the core of the manuscript.
However, in the case of cylindrical geometries, the 
(normalized) $g_v, \tilde{g}_v$ functions are obtained from Bessel's functions (with $x\rightarrow r$ the radius coordinate), and the situation is a bit more subtle \cite{AlexguideNJP}.
In particular, we have:
\begin{eqnarray}
\frac{\partial \tilde{g}_v(x)}{\partial x} & = & -k_c \, g_v(x) , \label{ouf} \\
\frac{\partial g_v(x)}{\partial x} & \neq & + k_c \, \tilde{g}_v(x) ,
\end{eqnarray}
the reason being that in the curved geometry, the operatorial form of the $x$-component is $1/x \, \partial(x \cdots )/\partial x$ instead of a "simple" derivative.  
This actually implies that   Eq. (\ref{pathos}) {\it cannot be interpreted as part of a gradient form} in a flat surface. But thankfully, Eq. (\ref{ouf}) is enough to guarantee that the charge-current vector is divergence-free, which is a physical requirement (charge conservation).
As well, the $g_v, \tilde{g}_v$ functions obtained in the cylindrical coordinate system {\it are not} periodic, and a $1/x$ pathology appeared in our integrals Eq. (\ref{scal2}); strictly speaking, {\it one cannot apply} our Hodge formalism as is. 

Interestingly, we can nonetheless follow the procedure developed in the core of the manuscript.
Take the curl expression Eqs. (\ref{Psix}-\ref{Psit}) and the load definition Eqs (\ref{loadx}-\ref{loads}), and simply replace the sine and cosine by the Bessel-derived $g_v, \tilde{g}_v$.
We end up again with relations of the type of Eqs. (\ref{eq1}-\ref{eq3}), introducing {\it virtual} operators $d_{c,x}^\dag, d_{c,z}^\dag, d_{c,t}^\dag$; the redundancy of the real electrode equation Eq. (\ref{redunt}) then brings that $d_{d,x}^\dag = d_{c,t}^\dag$, linking strictly the operators defined on the two distinct surfaces [and implying relations of the type of Eqs. (\ref{impose2}-\ref{CxCt})].
In this sense, the load operators appearing on the virtual plane are not so virtual, since they can be accessed from the real transverse current living on the real electrode.
From the transverse current continuity, we can pose in the cylindrical geometry:
\begin{equation}
L_x' = \frac{L_x}{|\tilde{g}_v(a)|} ,
\end{equation}
with $a$ the radius of the cylinder marking the end of the virtual surface. In the Cartesian case, $|\tilde{g}_v(\pm d/2)|=1$ for any $m>0$ and $L_x'=L_x$; in the cylinder geometry however, the normalization point $\tilde{g}_v(x_{max})=1$ {\it is not} on the border.
This finally defines the inductance of the real plane from the one calculated for the virtual one.
We can also perfectly well impose to the capacitances $C_z$ of virtual and real electrodes to be equal.
As well, this actually means that the $\varphi_{\beta}'$ real fields defined in Eq. (\ref{phiprime}) are equal to the virtual field's amplitudes of Eq. (\ref{gtilde}):
in this sense, {\it these can be considered as generalized fluxes}.
All coefficients involved in the modeling are summarized in Appendix \ref{curlparams}. \\ 

\begin{figure}[h!]
		\centering
	\includegraphics[width=11cm]{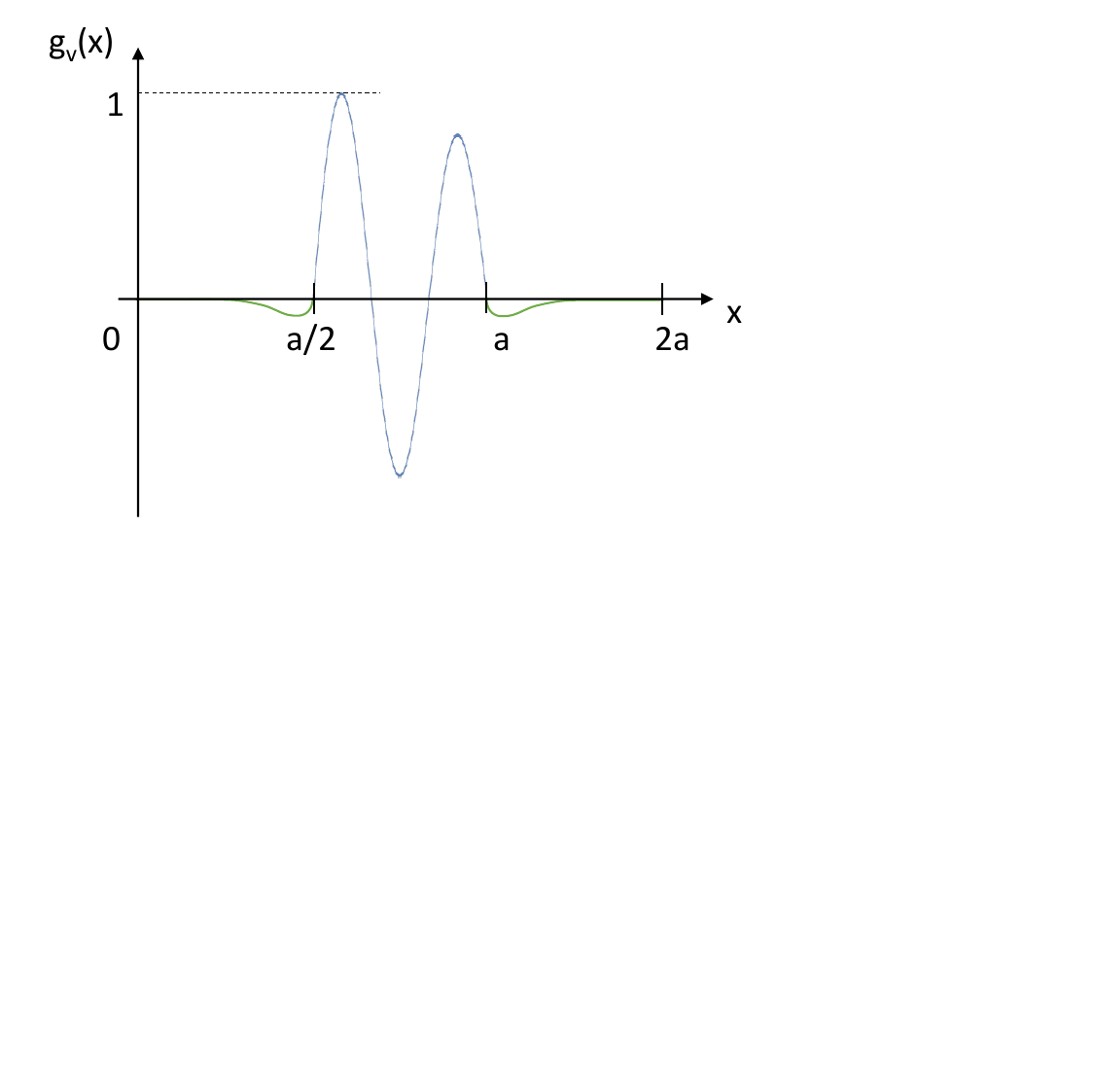}
           \vspace{-5.5cm}			
			\caption{ 
			 Illustration of the mathematical extension of the $g_v(x)$ transverse function for a coaxial mode. Here, explicitly we plot the $m=3$ case with $a=2 b$. The green line can be such that extremities are "ideally" flat around 0, with $|g_v(x)| \ll 1$ (see text).}
			\label{fig3}
\end{figure}

One therefore derives a scattering matrix here exactly like for the conventional modes. So, what is the matter with cylindrical TE$_{n=0,m}$ waves then?
To answer the question, let us {\it pose} a "generalized gradient" operator with an $x$-component given by $1/x \, \partial(x \cdots )/\partial x$, and keep our previous definitions for the curl and divergence. In the cylindrical geometry, $g_v(x)$ is meaningful on $[0,a]$ and is $\propto x$ near 0. For the coaxial case, if $b$ is the radius of the inner electrode we can arbitrarily extend in a $C^\infty$ fashion the function over $[0,b]$ such that  $d^n g_v(x=0)/d x^n=0$ 
with $n \geq 0$. We then apply the same "mathematical trick" as for the $\gamma(z)$ function in Section \ref{model}, with a $C^\infty$ extension of our $g_v$ functions over $[a,2a]$ such that $d^n g_v(x=2a)/d x^n=0$ for $n \geq 0$:
anti-symmetrysing $g_v(-x)=-g_v(x)$ and replicating it with a period $4a$, we have created a perfectly regular  and periodic (odd) function. We can therefore extend the integration limits of our bilinear forms to $[-2a,2a]$ and impose again periodic boundary conditions on all $x,z,t$, reminding that $g_v,\tilde{g}_v$ are physically meaningful only on a reduced domain (see Fig. \ref{fig3}).
Our Hodge-like properties read then:
\begin{eqnarray}
&& \!\!\!\!\!\!\!\!\!\!\!\!\!\!\!\!\!\!\!\!\!\!\! \langle \mbox{grad}(\phi),\vec{j} \,\rangle_{vector}  =  -\langle \frac{\phi}{x},j_x \rangle_{scalar}+\langle \phi,\mbox{div}(\vec{j})\, \rangle_{scalar} , \\
&& \!\!\!\!\!\!\!\!\!\!\!\!\!\!\!\!\!\!\!\!\!\!\! \mbox{curl} \left[ \mbox{grad}(\phi)\right]  =  \{ 0,-\frac{1}{ L_x L_z} \frac{\partial( \phi/x)}{\partial t'} \, ,+\frac{1}{ L_x L_z} \frac{\partial (\phi/x)}{\partial z}\} , \\
&& \Delta_{L,v}( \vec{j}\,) = -\frac{\mbox{div}( \vec{j}\,)/x}{L_x} + \Delta_{L}( \vec{j}\,) , 
\end{eqnarray}
with $\Delta_{L}$ our "conventional" Laplacian operator (see Section \ref{model}).
Since harmonic fields verify $\mbox{div}( \vec{j}_{harm})=0$, we can identify for them the two Laplacians. As well, the curl is still its own adjoint, and $div.curl=0$ identically. 
However, $curl.grad \neq 0$ in general, and gradient and divergence operators are {\it not} adjoint anymore;
 we shall not address any further here the case of the harmonic part. 

\begin{figure}[h!]
		\centering
	\includegraphics[width=9cm]{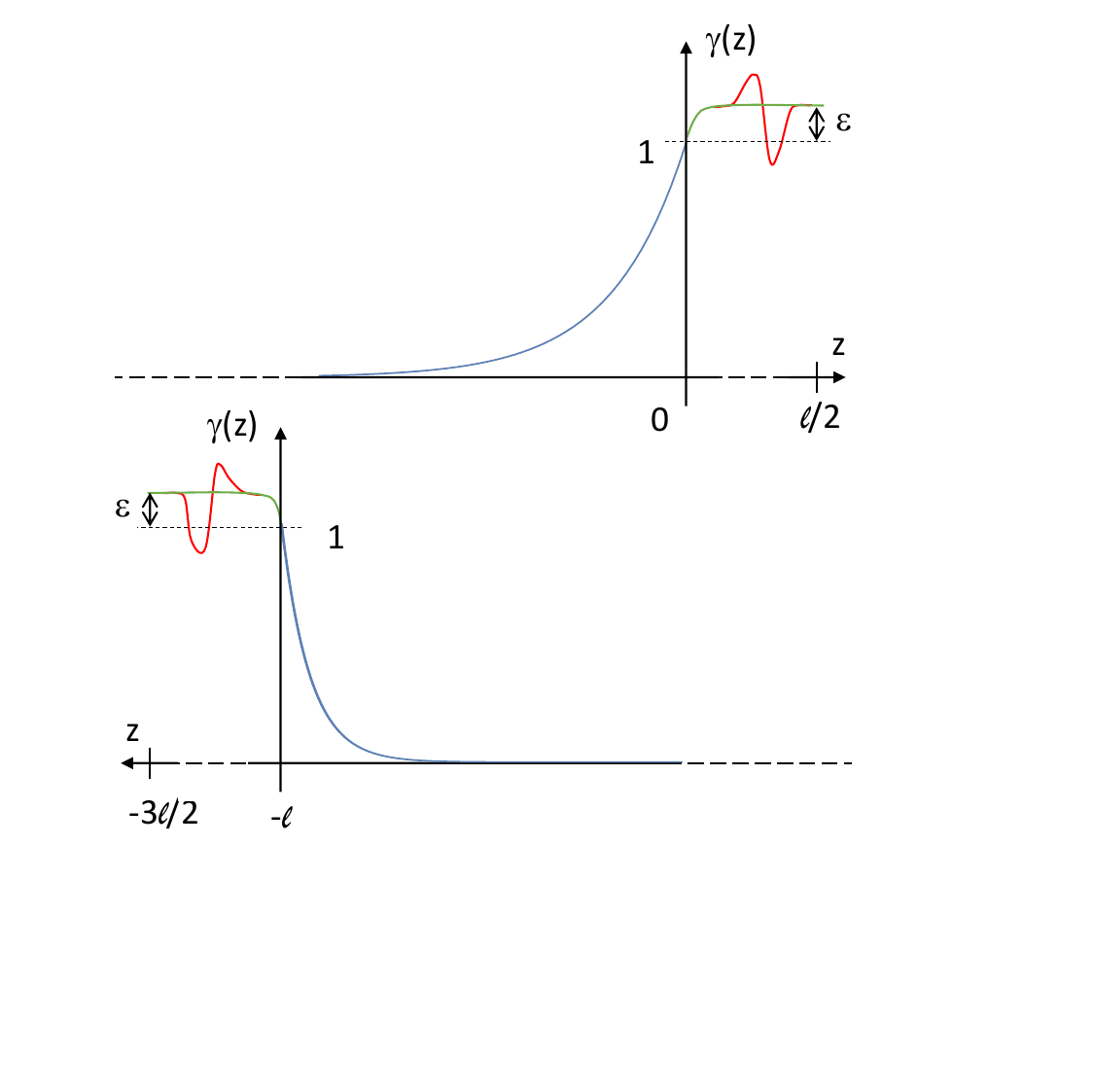}
           \vspace{-2.cm}			
			\caption{ 
			Illustration of the mathematical extension of the function $\gamma(z)$ [not to scale]. The two sides are in principle not symmetric, we simply require in the first place $\gamma(-3/2\ell)=\gamma(+\ell/2)$ and all derivatives equal to 0 at these extremities. The simplest possibility is the green curve. However, one can as well produce a more involved pattern (schematically the red one), fulfilling a specific integral property (see text). }
			\label{fig4}
\end{figure}

Interestingly, if we consider {\it only physical solutions} (meaning that $\phi$ corresponds to the traveling field $\varphi_{v,\alpha}$ and $\vec{j}$ to the load $\mbox{curl}[\vec{\Psi}_{v,\alpha}]$), we explicitly obtain:
\begin{eqnarray}
\langle \frac{\phi}{x},j_x \rangle_{scalar} & = & \mathbbm{i} \bar{v} \frac{\sqrt{2} \,\phi_{zpf}^2 \,T}{L_x} \int_{-2a}^{+2a} \frac{g_v(x)}{x} \tilde{g}_v(x) \, dx  \, \times \nonumber \\
&& \!\!\!\!\!\!\!\!\!\!\!\!\!\!\!\!\!\!\!\!\!\!\!\! \int_{-3\ell/2}^{+\ell/2} \left[ h_\phi(z) \, h_j(z)^\dagger + h_\phi(z)^\dagger \, h_j(z) \right] \, dz , \label{fuckstuff} \\
& \mbox{with:} & \nonumber \\
h_\phi(z) & =& -\mathbbm{i} c_{s,\beta}^\dagger \cos(\beta z) -  c_{a,\beta}^\dagger \sin(\beta z) , \\
h_j(z) & = & \frac{d \gamma(z)}{d z} \left[ c_{s,\beta}^\dagger \, A_{\alpha,s,t} + c_{a,\beta}^\dagger \, A_{\alpha,a,t} \right] \! ,
\end{eqnarray}
with the extended integrals introduced for $\gamma(z)$ in Section \ref{model}. 
There is no pathology in the $x$-integral when $x \rightarrow 0$. As well, one can easily comply with  
$ |g_v(x)| \ll 1$ on the extension domains, and 
$|d\gamma(z)/dz|<\kappa_\alpha$ or $\bar{\kappa}_\alpha$ with $|d\gamma(z)/dz| \sim 0$ almost everywhere on $[-3\ell/2,-\ell] \cup [0,+\ell/2]$ (see Figs. \ref{fig3} and \ref{fig4}, the green lines).
This means then that, under these conditions, the functions's extensions play no role in Eq. (\ref{fuckstuff}). 

We see that the integrals appearing in the above expression are:
\begin{eqnarray}
2k_c\!\! \int_{0 \, \mbox{or}\, b}^{a} \frac{g_v(x)}{x} \tilde{g}_v(x) \, dx & = & 3 \!\! \int_{0 \, \mbox{or}\, b}^{a} \left(g_v(x)/x\right)^2 \, dx, \label{firstfuck} \\
\int_{-\ell}^0 \frac{d \gamma(z)}{d z}  \cos (\beta z) \, dz &=&\!\!  +\frac{\kappa_\alpha^2}{\beta^2+\kappa_\alpha^2}-  \frac{\bar{\kappa}_\alpha^2}{\beta^2+\bar{\kappa}_\alpha^2}, \label{secfuck}  \\
\int_{-\ell}^0 \frac{d \gamma(z)}{d z}  \sin (\beta z) \, dz &=&\!\! -\frac{\beta\,\kappa_\alpha}{\beta^2+\kappa_\alpha^2}-  \frac{\beta\,\bar{\kappa}_\alpha}{\beta^2+\bar{\kappa}_\alpha^2} , \label{thirdfuck} 
\end{eqnarray}
in the limit $\kappa_\alpha \ell, \bar{\kappa}_\alpha \ell \gg 1$. Eq. (\ref{firstfuck}) is {\it strictly positive}.
Numerically, using formulas from Ref. \cite{AlexguideNJP} we find out that for a coaxial line it remains small, and even decreases with increasing mode number $m$ or $a/b \rightarrow 1^+$. However for a cylindrical hollow guide, it is not especially small. 
As well, while Eqs. (\ref{secfuck},\ref{thirdfuck}) are in absolute value smaller than 2, they {\it are nonzero}. 
But the point is that we can be more creative with our $\gamma(z)$ extension: we can perfectly well require that it {\it compensates} in Eq. (\ref{fuckstuff}) the contributions of Eqs. (\ref{secfuck},\ref{thirdfuck}), see the schematic (red line) in Fig. \ref{fig4}.
Under this condition, we recover our de Rham differential complex properties, but {\it only within the subspace corresponding to physical solutions} $\varphi_{\alpha,v}, \vec{\Psi}_{\alpha,v}$: $grad$ and $div$ are adjoint, and $\langle \mbox{curl}[\mbox{grad}(\varphi_{\alpha,v})],\vec{\Psi}_{\alpha,v} \,\rangle_{vector}  = 0$ replaces then the corresponding exact expression.
Which is actually all we look for here.
 
%

\section{Mode-dependent parameters}
\label{modeparams}

We remind in this Appendix the modal parameters, characteristic of each mode, that enable to compute all required quantities. We adapt here the expressions found in Refs. \cite{EddyguideNJP,AlexguideNJP}, in order to make them all compatible within the same framework: the one of the present manuscript. Incidentally, while being equivalent to the originals some expressions become more readable.
For the Cartesian geometries, the used formulas arise from Ref. \cite{EddyguideNJP}. Tab. \ref{tab_1} presents the parallel plate TEM and TM$_{m}$ modes, while Tab. \ref{tab_2}
lists TM$_{n,m}$ and TE$_{n>0,m}$ results in the case of a rectangular guide.
In Tab. \ref{tab_3}, the specific case of 
 TE$_{m}$ (parallel plate) and TE$_{n=0,m}$ (rectangular guide) waves is summarized. These are formally {\it equivalent}.
Also, the total capacitance $C_{tot}$ that enters into their $\phi_{zpf}$ calculation is given by  Eq. (\ref{Ctot}). 

\begin{table}[h!]
\center
\caption{Modal parameters for parallel plates "standard" waves. Electrode width $w$, gap $d$. }
\begin{tabular}{|c|c|} 
\hline
Wave type &   Parameters \\ \hline \hline
TEM    & $k_c=0$ \\
       & $h_{ef\!f}=d$ \\
       & $w_{ef\!f}=w$ \\ \hline
TM$_{m>0}$   & $k_c=k_{cy}$ with $k_{cy}=m\,\pi/d$ \\
           & $h_{ef\!f}=d/2$ \\
           & $w_{ef\!f}=w$ \\ \hline
\end{tabular} \label{tab_1}
\end{table}

\begin{table}[h!]
\center
\caption{Modal parameters for rectangular guide  "standard" waves. Electrode width $w$, gap $d$. }
\begin{tabular}{|c|c|} 
\hline
Wave type &   Parameters \\ \hline \hline
TM$_{n>0,m>0}$ & $k_c=\sqrt{k_{cx}^2+k_{cy}^2}$ with $k_{cx}=n\,\pi/w$, $k_{cy}=m\,\pi/d$ \\
           & $h_{ef\!f}=d/2 \left( \frac{k_c}{k_{cy}} \right)^2$ \\
           & $w_{ef\!f}=w/2$ \\ \hline
TE$_{n>0,m>0}$ & $k_c=\sqrt{k_{cx}^2+k_{cy}^2}$ with $k_{cx}=n\,\pi/w$, $k_{cy}=m\,\pi/d$ \\
           & $h_{ef\!f}=d/2 \left( \frac{k_c}{k_{cx}} \right)^2 $ \\
           & $w_{ef\!f}=w/2$ \\ \hline
\end{tabular} \label{tab_2}
\end{table}

\begin{table}[h!]
\center
\caption{Modal parameters for parallel plate and rectangular guide TE$_{n=0,m}$ waves. Electrode width $w$, gap $d$. }
\begin{tabular}{|c|c|} 
\hline
Wave type &   Parameters \\ \hline \hline
TE$_{m>0}$  & $k_c= k_{cy} $ with $k_{cy}=m\,\pi/d$ \\
or TE$_{n=0,m>0}$   & $h_{ef\!f}=w$ \\
                    & $w_{ef\!f}=d/2$ \\ \hline
\end{tabular} \label{tab_3}
\end{table}

The cylindrical geometry results are adapted from Ref. 
\cite{AlexguideNJP}.
$J_n$ and $Y_n$ are Bessel's functions of first and second kind respectively, of order $n$. The prime stands for their derivatives. The roots of the equations leading to $k_c$ are indexed by $m>0$.
In Tab. \ref{tab_4}, we present the parameters related to TEM, TM and TE$_{n\neq0,m}$ waves propagating in a coaxial line. In Tab. \ref{tab_6} are summarized the results applying to TM and TE$_{n\neq0,m}$ waves confined in a hollow cylinder. 
Finally the peculiar TE$_{n=0,m}$ waves characteristics are listed in Tab. \ref{tab_5} (coaxial) and Tab. \ref{tab_7} (cylinder). 
The $A_m(x_{max})$ parameter is a normalization amplitude required for virtual generalized fluxes; $x_{max}$ is defined as the radial position of the first $g_v$ maximum, the closest to $x=0$.
For the TE$_{n=0,m}$ modes, 
the total capacitance $C_{tot}$ that enters into the $\phi_{zpf}$ calculation is given by  Eq. (\ref{Ctot}). 

\begin{table}[h!]
\center
\caption{Modal parameters for the coaxial guide  "standard" waves. Outer radius $a$, inner radius $b$. }
\hspace*{-0.3cm}
\begin{tabular}{|c|c|} 
\hline
Wave type &   Parameters \\ \hline \hline
TEM        & $k_c =0 $   \\
           & $h_{ef\!f}=a \ln (a/b)$ \\
           & $w_{ef\!f}=2 \pi a$ \\ \hline
TM$_{n=0,m>0}$ & $k_c $ with $J_0(k_c a)Y_0(k_c b)=J_0(k_c b)Y_0(k_c a)$  \\
           & $h_{ef\!f}= a \left[ \frac{1}{2}-\frac{\pi^2 k_c^2 b^2}{8} \times \right. $  \\ 
           & $ \left. \left( J_1[k_c b]Y_0[k_c a]-J_0[k_c a]Y_1[k_c b] \right)^2 \right] $ \\
           & $w_{ef\!f}=2 \pi a$ \\ \hline
TM$_{n>0,m>0}$ & $k_c $ with $J_n(k_c a)Y_n(k_c b)=J_n(k_c b)Y_n(k_c a)$  \\
           & $h_{ef\!f}= a  \frac{\pi^2 k_c^2}{4} \times $ \\
           &$ \int_{b}^{a}\!  r  \left( J_n[k_c r] \, Y_n[k_c a] - 
  J_n[k_c a ] \, Y_n [ k_c r] \right)^2   dr$ \\
           & $w_{ef\!f}= \pi a$ \\ \hline
TE$_{n>0,m>0}$ & $k_c $ with $J_n'(k_c a)Y_n'(k_c b)=J_n'(k_c b)Y_n'(k_c a)$  \\
           & $h_{ef\!f}=a  \frac{k_c^2}{n^2}  \times $ \\
           & $ \!\!\!\!\!\!\!\!\!\!\!\!\!\!\!\!\!\!\!\!\!\!\!\!\!\!\!\!\!\!\!\!\!\! \int_{b}^{a} \! r \! \left( \!\frac{\left( J_{n+1}[k_c a]-J_{n-1}[k_c a]\right) \, Y_n[k_c r] - 
  J_n[k_c r ] \, \left( Y_{n+1} [ k_c a]-Y_{n-1} [ k_c a] \right) }{ \left( J_{n+1}[k_c a]-J_{n-1}[k_c a]\right) \, Y_n[k_c a] - 
  J_n[k_c a ] \, \left( Y_{n+1} [ k_c a]-Y_{n-1} [ k_c a] \right)} \!\right)^{\!\!2} \!\!  dr$  \\
           & $w_{ef\!f}=\pi a$ \\ \hline
\end{tabular} \label{tab_4}
\end{table}

\begin{table}[h!]
\center
\caption{Modal parameters for the hollow cylinder guide  "standard" waves. Radius $a$. }
\begin{tabular}{|c|c|} 
\hline
Wave type &   Parameters \\ \hline \hline
TM$_{n=0,m>0}$ & $k_c $ with  $J_0(k_c a)=0$ \\
           & $h_{ef\!f}=a$ \\
           & $w_{ef\!f}=\pi a$ \\ \hline
TM$_{n>0,m>0}$ & $k_c $ with $J_n(k_c a)=0$  \\
           & $h_{ef\!f}=a  \left(\frac{2 J_{n-1} [k_c a]}{J_{n-1} [k_c a]-J_{n+1} [k_c a]}\right)^2  $ \\
           & $w_{ef\!f}=\pi a/2$ \\ \hline
TE$_{n>0,m>0}$ & $k_c $ with $J_{n-1}(k_c a)=J_{n+1}(k_c a)$  \\
           & $h_{ef\!f}=a \left( \frac{k_c }{k_{cx} }\right)^{\!\!2}    \left(1- \frac{  2 n J_{n+1}[k_c a]}{ (k_c a)\, J_n[k_c a]} +\frac{  J_{n+1}[k_c a]^2 }{  J_n[k_c a]^2 } \right)$ \\
           & $w_{ef\!f}=\pi a/2$ \\ \hline
\end{tabular} \label{tab_6}
\end{table}

\begin{table}[h!]
\center
\caption{Modal parameters for the coaxial guide  TE$_{n=0,m}$ waves. Outer radius $a$, inner radius $b$; $A_m(x_{max})$ is profile amplitude normalization (see text). }
\begin{tabular}{|c|c|} 
\hline
Wave type &   Parameters \\ \hline \hline
TE$_{n=0,m>0}$ & $k_c $ with $J_1(k_c a)Y_1(k_c b)=J_1(k_c b)Y_1(k_c a)$  \\
           & $h_{ef\!f}=a \frac{8-2 \left( k_c b\right)^2 \pi^2 \left( J_1[k_c a] Y_0[k_c b] - J_0[k_c b]Y_1[k_c a] \right)^2}{\left( k_c a \right)^4 \pi^2 A_m^2} $ \\
           & $A_m(x_{max})=2\frac{J_1[k_c a]Y_1[k_c x_{max}] - J_1[k_c x_{max}] Y_1[k_c a]}{k_c a}$ \\
           & with $\frac{J_0(k_c x_{max})-J_2(k_c x_{max})}{Y_0(k_c x_{max})-Y_2(k_c x_{max})}=\frac{J_1(k_c a)}{Y_1(k_c a)} $ \\ 
           & $w_{ef\!f}=2 \pi a$ \\ \hline
\end{tabular} \label{tab_5}
\end{table}

\begin{table}[h!]
\center
\caption{Modal parameters for the hollow cylinder guide  TE$_{n=0,m}$ waves. Radius $a$; $A_m(x_{max})$ is profile amplitude normalization (see text). }
\hspace*{-0.3cm}
\begin{tabular}{|c|c|} 
\hline
Wave type &   Parameters \\ \hline \hline
TE$_{n=0,m>0}$ & $k_c $ with $J_{1}(k_c a)=0$  \\
           & $h_{ef\!f}=  a \left( 2 \frac{J_0[k_c a]}{A_m} \right)^2 $  \\
           & $A_m = 2 J_1(k_c x_{max})$ with $J_{0}(k_c x_{max})=J_{2}(k_c x_{max})$ \\
           & $w_{ef\!f}=\pi a$ \\ \hline
\end{tabular} \label{tab_7}
\end{table}


\section{Curl coefficients}
\label{curlparams}

We present in this Appendix the coefficients obtained when solving the problem at hand. We remind that the curl coefficients {\it are not } unique, while the final result is. We made here choices which presumably make the reading more tractable. 
In Tab. \ref{tab_8}, we list the results applying to TEM and TM$_{n=0,m}$ waves. Tab. \ref{tab_9} summarizes the case of TM$_{n>0,m}$ modes, and Tab. \ref{tab_10} the situation resulting from TE$_{n>0,m}$ ones.
The specific case of TE$_{n=0,m}$ waves is treated in Tab. \ref{tab_11}. We give there the coefficients applying to the {\it real} electrode, and therefore the scattering matrix redundant parameters ${\cal K}_{\alpha,i,x}$ (with $i=s,a$). These are equal by construction to ${\cal K}_{\alpha,s,t}$ and ${\cal K}_{\alpha,a,t}$ respectively, computed for virtual fields. The {\it virtual} electrode coefficients are the same as the ones of Tab. \ref{tab_10}, but for a curl expression written in the virtual plane (see Appendix \ref{topomodes}).

\begin{table}[h!]
\center
\caption{Curl coefficients and scattering matrix coefficients obtained for TEM and TM$_{n=0,m}$ waves. }
\begin{tabular}{|c|c|} 
\hline
coefficient &   value \\ \hline \hline
$A_{\alpha,s,x}$ & to be found  \\
$A_{\alpha,a,x}$ & to be found  \\
$B_{\alpha,s,x}$ & 0  \\
$B_{\alpha,a,x}$ & 0  \\
$A_{\alpha,s,z}$ & 0  \\
$A_{\alpha,a,z}$ & 0  \\
$B_{\alpha,s,z}$ & 0  \\ 
$B_{\alpha,a,z}$ & 0  \\ 
$A_{\alpha,s,t}$ & 0  \\
$A_{\alpha,a,t}$ & 0  \\
$B_{\alpha,s,t}$ & 0  \\
$B_{\alpha,a,t}$ & 0  \\  \hline
    ${\cal K}_{\alpha,s,z}$      & $ A_{\alpha,s,x}$ \\
    ${\cal K}_{\alpha,a,z}$      & $ A_{\alpha,a,x}$ \\ \hline
\end{tabular} \label{tab_8}
\end{table}

\begin{table}[h!]
\center
\caption{Curl coefficients and scattering matrix coefficients obtained for TM$_{n>0,m}$ waves. }
\begin{tabular}{|c|c|} 
\hline
coefficient &   value \\ \hline \hline
$A_{\alpha,s,x}$ & 0  \\
$A_{\alpha,a,x}$ & 0  \\
$B_{\alpha,s,x}$ & to be found  \\
$B_{\alpha,a,x}$ & to be found  \\
$A_{\alpha,s,z}$ & 0  \\
$A_{\alpha,a,z}$ & 0  \\
$B_{\alpha,s,z}$ & 0  \\ 
$B_{\alpha,a,z}$ & 0  \\ 
$A_{\alpha,s,t}$ & 0  \\
$A_{\alpha,a,t}$ & 0  \\
$B_{\alpha,s,t}$ & 0  \\
$B_{\alpha,a,t}$ & 0  \\  \hline
    ${\cal K}_{\alpha,s,z}$      & $B_{\alpha,s,x}$ \\
    ${\cal K}_{\alpha,a,z}$      & $B_{\alpha,a,x}$ \\ \hline
\end{tabular} \label{tab_9}
\end{table}

For each configuration, {\it only 2 variables are free and must be found}. These are obtained from the load impedance parameters $r_z$ and $\varphi_z$. Solving for a scattering matrix leads to, making use of Eqs. (\ref{kacal1}-\ref{kacal2}):
\begin{eqnarray}
&& \!\!\!\!\!\!\!\!\!\!\!\!\!\!\!\!\!\!\!\! \bar{{\cal K}}_{\alpha,z}  =  \frac{\left(r_z^2-\mathbbm{i} \eta  \right)\left(\frac{1}{\cos(\varphi_z)}-1 \right)-r_z \left(\eta -\mathbbm{i} \right) \tan(\varphi_z) }{r_z^2+\eta^2} , \\
&&\!\!\!\!\!\!\!\!\!\!\!\!\!\!\!\!\!\!\!\!\!\!\!\!\! \Delta{\cal K}_{\alpha,z}  =  \frac{\left(r_z^2+\mathbbm{i} \eta  \right)\left(\frac{1}{\cos(\varphi_z)}-1 \right)-r_z \left(\eta +\mathbbm{i} \right) \tan(\varphi_z)}{r_z^2+\eta^2} . 
\end{eqnarray}
Injecting these in the tabulars above, one obtains the sought curl coefficients.
Remarkably, only if $\varphi_z=0$ (purely {\it real load}) do these expressions equal 0 simultaneously. For all waves, this actually leads to a zero curl vector.
However for $\varphi_z \neq 0$, the curl contribution is strictly nonzero.

\begin{table}[h!]
\center
\caption{Curl coefficients and scattering matrix coefficients obtained for TE$_{n>0,m}$ waves. Equivalently, virtual electrode decomposition of TE$_{n=0,m}$ modes (see Appendix \ref{topomodes}).}
\begin{tabular}{|c|c|} 
\hline
coefficient &   value \\ \hline \hline
$A_{\alpha,s,x}$ & 0  \\
$A_{\alpha,a,x}$ & 0  \\
$B_{\alpha,s,x}$ & $-A_{\alpha,s,t}\frac{k}{k_{cx}}$  \\
$B_{\alpha,a,x}$ & $-A_{\alpha,a,t}\frac{k}{k_{cx}}$  \\
$A_{\alpha,s,z}$ & 0  \\
$A_{\alpha,a,z}$ & 0  \\
$B_{\alpha,s,z}$ & 0  \\ 
$B_{\alpha,a,z}$ & 0  \\ 
$A_{\alpha,s,t}$ & to be found  \\
$A_{\alpha,a,t}$ & to be found  \\
$B_{\alpha,s,t}$ & 0  \\
$B_{\alpha,a,t}$ & 0  \\  \hline
    ${\cal K}_{\alpha,s,z}$      & $-A_{\alpha,s,t}\frac{\beta}{k_{cx}}$ \\
    ${\cal K}_{\alpha,a,z}$      & $-A_{\alpha,a,t}\frac{\beta}{k_{cx}}$ \\ \hline
\end{tabular} \label{tab_10}
\end{table}

\begin{table}[h!]
\center
\caption{Curl coefficients and scattering matrix coefficients obtained for TE$_{n=0,m}$ waves ({\it real} electrode related). The ${\cal K}_{\alpha,i,x}$ are equal to the ${\cal K}_{\alpha,i,t}$ of the virtual electrode, see Appendix \ref{topomodes}.}
\begin{tabular}{|c|c|} 
\hline
coefficient &   value \\ \hline \hline
$A_{\alpha,s,x}$ & 0  \\
$A_{\alpha,a,x}$ & 0  \\
$B_{\alpha,s,x}$ & 0  \\
$B_{\alpha,a,x}$ & 0  \\
$A_{\alpha,s,z}$ & to be found  \\
$A_{\alpha,a,z}$ & to be found  \\
$B_{\alpha,s,z}$ & 0  \\ 
$B_{\alpha,a,z}$ & 0  \\ 
$A_{\alpha,s,t}$ & 0  \\
$A_{\alpha,a,t}$ & 0  \\
$B_{\alpha,s,t}$ & 0  \\
$B_{\alpha,a,t}$ & 0  \\  \hline
    ${\cal K}_{\alpha,s,x}$      & $\mathbbm{i} \,A_{\alpha,s,z} \frac{k}{ k_c (L_x/L_x')}$  \\
    ${\cal K}_{\alpha,a,x}$      & $\mathbbm{i} \,A_{\alpha,a,z} \frac{k}{ k_c (L_x/L_x')}$ \\ \hline
\end{tabular} \label{tab_11}
\end{table}


\newpage

\begin{thebibliography}{blaaaaaaaaaaaaaaaaaaaaaaaaaaaaaaaaaaaaaaaaaaaa}
\bibitem{qubit} Matthew P. Bland, Faranak Bahrami, Jeronimo G. C. Martinez, Paal H. Prestegaard, Basil M. Smitham, Atharv Joshi, Elizabeth Hedrick, Shashwat Kumar, Ambrose Yang, Alexander C. Pakpour-Tabrizi, Apoorv Jindal, Ray D. Chang, Guangming Cheng, Nan Yao, Robert J. Cava, Nathalie P. de Leon and Andrew A. Houck, 
{\it Millisecond lifetimes and coherence times in 2D transmon qubits}, Nature {\bf 647}, pages 343–348 (2025).
\bibitem{opto} G.A. Peterson, F. Lecocq, K. Cicak, R.W. Simmonds, J. Aumentado, and J.D. Teufel, {\it Demonstration of Efficient Nonreciprocity in a Microwave Optomechanical Circuit}, Phys. Rev. X {\bf 7}, 031001 (2017).
\bibitem{Twpa} Luca Planat, Arpit Ranadive, Remy Dassonneville, Javier Puertas Martinez, Sebastien Leger, Cecile Naud, Olivier Buisson, Wiebke Hasch-Guichard, Denis M. Basko, Nicolas Roch, {\it A photonic crystal Josephson traveling wave parametric amplifier}, Phys. Rev. X {\bf 10}, 021021 (2020).
\bibitem{loopholewalraff}  Simon Storz, Josua Sch\"ar, Anatoly Kulikov, Paul Magnard, Philipp Kurpiers,
Janis L\"utolf, Theo Walter, Adrian Copetudo, Kevin Reuer, Abdulkadir Akin,
Jean-Claude Besse, Mihai Gabureac, Graham J. Norris, Andr\'es Rosario, Ferran Martin,
Jos\'e Martinez, Waldimar Amaya, Morgan W. Mitchell, Carlos Abellan, Jean-Daniel Bancal,
Nicolas Sangouard, Baptiste Royer, Alexandre Blais and Andreas Wallraff, {\it Loophole-free Bell inequality violation with superconducting circuits}, Nature Vol. 617, p. 265 (2023).
\bibitem{gardiner} C.W. Gardiner, P. Zoller, {\it Quantum Noise}, Springer Series in Synergetics, Third Ed. (2004).
\bibitem{quantumpulses} Alexander Holm Kiilerich and Klaus M\o lmer, {\it Input-Output Theory with Quantum Pulses}, Phys. Rev. Lett. {\bf 123}, 123604 (2019).
\bibitem{molmer} A.H. Kiilerich, K. M\o lmer, {\it Quantum interactions with pulses of radiation}, 
Physical Review A {\bf 102} (2), 023717 (2020).
\bibitem{clerk} A.A. Clerk, {\it Quantum-limited position detection and amplification: A linear response perspective}, Phys. Rev. B {\bf 70}, 245306 (2004).
\bibitem{walraff}  Alexandre Blais, Arne L. Grimsmo, S. M. Girvin, Andreas Wallraff, {\it Circuit Quantum Electrodynamics}, Rev. Mod. Phys. {\bf 93}, 025005 (2021).
\bibitem{blencowe} P. D. Nation, J. R. Johansson, M. P. Blencowe, Franco Nori, {\it Stimulating Uncertainty: Amplifying the Quantum Vacuum with
Superconducting Circuits}, Rev. Mod. Phys. Vol. 84, 1-27 (2012).
\bibitem{nori} Xiu Gu, Anton Frisk Kockum, Adam Miranowicz , Yu-xi Liu, Franco Nori, {\it Microwave photonics with superconducting quantum circuits},  Physics Reports Vols. 718–719,pp. 1–102 (2017).
\bibitem{devoretRMP} A.A. Clerk, M.H. Devoret,
S.M. Girvin, Florian Marquardt, and R.J. Schoelkopf, {\it Introduction to Quantum Noise, Measurement and Amplification}, Rev. Mod. Phys. {\bf 82}, 1155 (2010).
\bibitem{pozar} David M. Pozar, {\it Microwave Engineering}, John Wiley \& Sons Inc., Fourth Ed. (2012). 
\bibitem{devoret} Michel Devoret, in {\it Quantum Fluctuations} (Les Houches Session LXIII) (Elsevier, Amsterdam), pp. 351-86. (1997).
\bibitem{frankel} Theodore Frankel, {\it The geometry of physics: An Introduction}, Cambridge University Press,  2nd ed. (2004).
\bibitem{EddyguideNJP} E. Collin, A. Delattre, {\it Waveguides in a quantum perspective}, New J. Phys. {\bf 27}, 093502 (2025); Corrigendum: New J. Phys. {\bf 28}, 059501 (2026).
\bibitem{AlexguideNJP} A. Delattre, E. Collin, {\it Canonical Quantization of Cylindrical Waveguides: A Gauge-Based Approach}, J. Phys. Commun. {\bf 10}, 075004 (2026).
\bibitem{leggett} A. Caldeira and A. Leggett, {\it Quantum Tunneling in a Dissipative System}, Ann. Phys. {\bf 149}, 374-456 (1983).

\end{thebibliography}
\end{document}